\documentclass{aa}
\pdfoutput=1
\usepackage[utf8]{inputenc}
\usepackage{graphicx}
\usepackage{natbib}
\usepackage{booktabs,tablefootnote}
\usepackage{threeparttable}
\usepackage[usenames, dvipsnames]{color}
\usepackage{subfigure}
\usepackage{enumitem}

\usepackage{txfonts}

\usepackage{amstext}

\begin{document}

   \title{Red-skewed K$\alpha$ iron lines in GX 13+1.}

   \author{T. Maiolino,
              \inst{1} 
              P. Laurent,
              \inst{2,3}
              L. Titarchuk,
              \inst{1}
              M. Orlandini,
              \inst{4}
              F. Frontera.
              \inst{1,4,5}  
          }

   \institute{University of Ferrara, Department of Physics and Earth Science,
             Via Saragat 1, Cap 44122, Ferrara, Italy\\           
            \email{maiolino@fe.infn.it}
         \and 
              CEA Saclay, 91191 Gif sur Yvette, France
         \and
              Laboratoire APC, 10 rue Alice Domont et Leonie Duquet, 75205 Paris Cedex 13, France  
         \and
              INAF/Osservatorio di Astrofisica e Scienza dello Spazio (OAS) Bologna, Via Gobetti 101, 40129 Bologna, Italy    
         \and
              ICRANET Piazzale d. Repubblica 10-12 65122 Pescara (PE), Italy
    }

   \date{}

  \abstract
   {Broad, asymmetric, and red-skewed Fe K$_\alpha$ emission lines have been observed in the spectra of low-mass X-ray binaries hosting neutron stars (NSs) as a compact object. Because more than one model is able to describe these features, the explanation of where and how the red-skewed Fe lines are produced is still a matter of discussion. It is broadly accepted that the shape of the Fe K$_\alpha$ line is strongly determined by the special and general relativistic effects occurring in the innermost part of the accretion disk. In this relativistic framework, the Fe fluorescent lines are produced in the innermost part of the accretion disk by reflection of hard X-ray photons coming from the central source (corona and/or NS surface). We developed an alternative and nonrelativistic model, called the \textsc{windline} model, that is capable to describe the Fe line features. In this nonrelativistic framework, the line photons are produced at the bottom of a partly ionized outflow (wind) shell as a result of illumination by the continuum photons coming from the central source. In this model the red-skewness of the line profile is explained by repeated electron scattering of the photons in a diverging outflow.
   }
   {Examining the asymmetry of the fluorescent Fe K emission line evident in the XMM-Newton EPIC-pn spectra of the NS source GX~13+1, we aim to distinguish between the two line models. Because GX~13+1 is a well-known disk-wind source, it is a perfect target for testing the \textsc{windline} model and compare the spectral fits between the relativistic and nonrelativistic line models. 
   }
   {We used two XMM-Newton EPIC-pn observations in which the Fe line profiles were previously reported in the literature. These observations are not strongly affected by pile-up, and the Fe emission lines appear very strong and red-skewed. In order to access the goodness of the fit and distinguish between the two line models, we used the run-test statistical method in addition to the canonical $\chi^2$ statistical method. 
   }
   { The \textsc{diskline} and \textsc{windline} models both fit the asymmetric Fe line well that is present in the XMM-Newton energy spectra of GX~13+1.
  From a statistical point of view, for the two observations we analyzed, the run-test was not able to distinguish between the two Fe line models, at 5\% significance level.
   } 
      {}


   \maketitle
%

\section{Introduction}
\label{sec:introduction}

The strong Fe $K_\alpha$ emission line (in the $\sim6.4-7.0$~keV X-ray energy band) has been observed as broadened, asymmetric, and red-skewed in a number of sources, which include extragalactic and Galactic accreting X-ray compact objects: Seyfert-type I active galactic nuclei (AGNs) \citep[e.g.,][]{Tanaka1995b,Nandra1997,Fabian2000, Reynolds2003,Miller2007,Hagino2016}, black hole (BH) low-mass X-ray binaries (LMXBs) \citep[e.g.,][]{Fabian1989,Miller2002b,Miller2007,Reynolds2003}, and neutron star (NS) LMXBs \citep[e.g.,][]{Bhattacharyya2007,Cackett2008,Pandel2008,Shaposhnikov2009,Reis2009,Cackett2010,Ludlam2017,LudlamXTE2017}. 

For accreting NSs and BHs, the standard interpretation for the Fe line asymmetry invokes general relativity (GR) effects through the strong gravitational field that is present close to the compact object, where the emission line is assumed to be produced. Although the relativistic interpretation is commonly accepted to explain the asymmetry of the observed iron line profiles, there are some considerations before this prevailing scenario can be accepted:

\begin{enumerate}
\item {At the time of writing, the relativistic line interpretation has some difficulties to explain all the characteristics that are observed in these sources (e.g., X-ray variability, see section \ref{sec:NSline}).}
\item {Fe line profiles with similar characteristics (i.e., broadened and red-skewed) were observed in the X-ray spectra of accreting white dwarfs (WDs) \citep[][Maiolino et al. in preparation]{Hellier2004,Vrielmann2005,Titarchuk2009}. These systems require an alternative explanation for the line profile because GR effects do not play a role around WDs.}
\item {A line model based on relativistic effects is not unique in describing the broad and asymmetric Fe K emission line profiles \citep[e.g.,][]{Hagino2016}}.
\end{enumerate}   

An alternative model, called the wind line model (\textsc{windline}, hereafter; presented by \citet{Laurent2007}), is able to explain the red-skewed Fe emission line profile observed in NS and BH LMXBs \citep{Titarchuk2009,Shaposhnikov2009}, and in systems containing WD as the accreting compact object (i.e., cataclysmic variables) because the line model does not require GR effects  \citep{Titarchuk2009}. In the \textsc{windline} framework, the asymmetric and red-skewed line profile is produced by repeated down-scattering of the line photons by electrons in a diverging outflow (wind). The line is generated in a partially ionized and thin inner shell of the outflow by irradiation of hard X-ray photons coming from the central source \citep[for more details about the model, see][and discussion therein]{Laurent2007}.

Motivated by the potential of the \textsc{windline} model in describing red-skewed iron line profiles in all three types of accreting X-ray powered sources, we applied this model on the residual excess in the Fe K energy range. This excess is observed in the spectra of all types of Galactic accreting compact objects: NSs, BHs, and WDs. In this first paper we present the investigation of the iron emission line profile in the spectrum of the LMXB GX~13+1, which hosts a neutron star as a compact object. 

GX 13+1 is a perfect target for testing the \textsc{windline} model because strong red-skewed Fe emission lines are reported in its spectra \citep[e.g.,][]{Trigo2012}, with simultaneous observations of blueshifted Fe absorption lines. This indicates an outflow around the source \citep{Boirin2005}, (see section \ref{sec:gx13p1}).

This paper is structured as follows: in the section \ref{sec:NSline} we present a concise background on broad and red-skewed iron lines in accreting NSs and their relevance. In section \ref{sec:windXSPEC} we describe the parameters of the \textsc{windline} and in \ref{sec:gx13p1} the characteristics of the source GX 13+1. In section \ref{sec:data} we show the XMM-Newton observations that we analyzed and the data reduction. In section \ref{sec:datanalysis} we describe the data analysis and compare the \textsc{windline} and the relativistic \textsc{diskline} fits. We discuss our results and conclude in section \ref{sec:dis}.

\subsection{Broad and red-skewed iron lines in accreting neutron stars}
\label{sec:NSline}

Fluorescent Fe K emission line profiles were first observed as broadened, asymmetric, and red-skewed in Galactic and extragalactic accreting BHs; in LMXBs and Seyfert-type I galaxies   \citep[e.g.,][and references therein]{Fabian1989,Tanaka1995b,Miller2007}. Because the Fe K band was successfully modeled by a relativistic line model, this led to the conclusion that the breadth and asymmetry of the line in BH sources is due to effects of Doppler-broadening, transverse Doppler shift, relativistic beaming, and gravitational redshifts produced in the innermost part of the accretion disk, where GR effects are playing a role. 

The subsequent discovery of red-skewed Fe emission line in the NS LMXB Serpens X--1 \citep{Bhattacharyya2007, Cackett2008}, followed by the observation in a few more NS LMXB sources, and the satisfactory fits of the line profiles using the relativistic line model extended the interpretation of the asymmetric line profiles in terms of GR effects on accreting NSs.

In the framework of the relativistic models, the fluorescent Fe emission line is produced by the reflection of hard photons (described by a power-law spectrum, coming from the inner part of the source) by the iron (ions) on the surface of a cold ($<$ 1 keV) accretion disk. 
The relativistic scenario is even now commonly accepted as an explanation for red-skewed Fe K emission lines in NS (and BH) sources. The line in NS LMXBs is usually modeled assuming a Schwarzschild potential around the accreting compact object (\textsc{diskline} model \citep{Fabian1989}). However, models considering a Kerr metric, created primarily for describing the emission line profiles coming from the accretion disks around rotating BHs, have also been used to fit to the iron emission lines in accreting NSs, see, for example, the \textsc{laor} \citep{Laor1991}, \textsc{relconv} \citep{Dauser2010}, and \textsc{relxill} \citep{Garzia2014} models. The main motivation for applying the relativistic scenario to NS sources is the fact that the inner disk radii predicted by the relativistic model have been found to agree well with the radii implied by kilohertz quasi-periodic oscillation (kHz QPO) frequency in 4U 1820-30 and GX394+2 sources as a Keplerian frequency at those radii (supporting the inner disk origin for kHz QPOs) \citep{Cackett2008}.

However, \citet{Shaposhnikov2009} and \citet{Titarchuk2009}, using \textit{Suzaku} and \textit{XMM-Newton} data, examined the red-skewed line profile observed in the spectra of two NS LMXB sources, Cyg~X--2 and Serpens X-1, in the framework of the \textsc{windline} model. They found that this nonrelativistic model is able to reproduce the red-skewed line profile with a fit quality (on the basis of $\chi^2$-statistic) similar to that obtained by relativistic reflection models. Although they were unable to conclusively rule out one of the models, they pointed out that the \textsc{windline} model appears to give a more adequate explanation because it does not require the accretion disk inner edge to advance close to the NS surface. In addition, the \textsc{windline} model explains a timing source property that the \textsc{diskline} model has difficulty to account for. They verified a lack of erratic fast variability when the Fe line is present in the Cyg~X--2 spectrum (which was also observed in a BH LMXB source, GX~339$-$4; \citep{Titarchuk2009}). This result weakens the red-skewed line connection with kHz QPOs and strengthens its connection to outflow phenomena because the suppression of fast variability can be explained by smearing in a strong outflow (wind) \citep[see Figure~8, Sections~3.1 and 3.2 in][]{Titarchuk2009}.

Furthermore, \citet{Lyu2014} used \textit{Suzaku}, \textit{XMM-Newton}, and \textit{RXTE} observations of the NS 4U 1636-53 to study the correlation between the inner disk radius (obtained from the Fe line relativistic fits) and the flux. They observed that the line did not change significantly with flux states of the source and concluded that the line is broadened by mechanisms other than just relativistic broadening. This behavior is not only observed in NS LMXBs. For example, observations of MCG--6-30-15, NGC~4051, and MCG--5-23-16 \citep{Reynolds1997,Reeves2006,Marinucci2014} show that the broad and red-skewed Fe lines in AGNs show little variability despite the large changes in the continuum flux. The iron line and the reflected emission flux do not respond to the X-ray continuum level of variability on short timescales \citep{Reeves2006}, which means that the fast changes that are almost certain to occur near the compact object are not observed in the line. This point is difficult to explain by the reflection and GR interpretation of the iron line formation, and has been suggested to be caused by strong gravitational light bending \citep[see][]{Miniutti2004,Chiang2011}.

In addition, \citet{Mizumoto2018}  demonstrated that the short lag time of the reprocessed Fe-K line energy band observed in AGNs can also be explained by scattering of photons coming from the X-ray central source by an outflow or disk-wind (with velocity of $\sim$~0.1c, placed at large distances, $\sim$ 100~R$_g$, from the compact object). Their simulations  resulted in features in the lag-energy plot and characteristics similar to those observed in 1H0707-495 and NGC 4151. They showed that the observation of short lag times does not necessarily indicate that the line is produced very close to the event horizon of the BH, as expected by the relativistic scenario  \citep[see][and references therein]{Mizumoto2018}.

As previously mentioned, the prevailing interpretation that has been used to model the red-skewed Fe line profiles in accreting NS (and also BHs) with reflection and GR effects is not entirely successful, and there is the alternative \textsc{windline} model that does not need relativistic effects, but successfully describes the observed line profiles in these sources. The importance to make this comparative study between the models is not just to either confirm or reject the relativistic framework in NS LMXBs, but also to confirm or reject the implications to the NS physics derived by the  models. By interpreting the broadening and red-skew of the Fe K emission line as due to GR effects, the relativistic models allow us to derive directly from the spectral fit parameters such as disk inclination, spin of the compact object, and inner radius of the accretion disk. The inner radius has been used to set an upper limit on the magnetic field strength and on the radius of the neutron stars, which in turn is used to study constraints on the NS equation of state \citep{Ludlam2017}. Although the relativistic models allow this powerful derivation, it is important to study the asymmetry of the Fe K line with other models that provide an alternative as well as a more consistent physical scenario of the NS LMXBs spectral emission.

\subsection{\textsc{windline} model into XSPEC} 
\label{sec:windXSPEC}

The main obstacle to analyzing the spectral data with the \textsc{windline} model is that the analytical solution is not available \citep{Shaposhnikov2009}. Therefore, the \textsc{windline} model was inserted within XSPEC with a number of FITS tables containing the results of the wind line model implemented using the Monte Carlo approach \citep{Laurent2007}. Each FITS table is constructed for a fixed emission line energy, running a grid of values for the free parameters. The free parameters of the model are as follows: 

\begin{itemize}[label={$\bullet$}]
\item the optical depth of the wind/outflow ($\tau_w$); 

\item the temperature of the electrons in the wind/outflow ($kT_{ew}$ in keV); 

\item the velocity ($\beta$) of the wind/outflow in unity of $c$ (speed of light); 

\item the redshift of the source; \textit{\textup{and}} 

\item the normalization (which is the number of photons in the whole spectrum, in units of photon/keV).
\end{itemize}

In this model, the line profiles strongly depend on the optical depth $\tau_w$ and on the velocity $\beta$ (v/c) of the outflow (wind). The shape of the red wing, below the broad peak, follows a power law with an index that is a strong function of $\tau_w$ and $\beta$. If the temperature of the electrons in the wind is low, kT$_{ew} < 1$ keV, the temperature does not change the line profile much, and most of the effect is due to the wind velocity, that is, the velocity effect is strong enough to change the red wing of the line at the observed level.

\subsection{Source GX 13+1}
\label{sec:gx13p1}
The source GX 13+1 is classified as a LMXB with a NS as a compact object. In LMXBs the secondary star (which is an evolved late-type K5 III giant star in GX 13+1) fills the Roche lobe and transfers mass onto the compact object through the inner Lagrangian point (L1), feeding the primary star through an accreting disk \citep{Lewin2006}.

This source is located at a distance of $7 \pm 1$ kpc and is also classified as a type I X-ray burster \citep{Fleischman1985, Matsuba1995}. Its bright persistent X-ray emission shows characteristics of both atoll and Z LMXBs \citep[see, e.g.,][and references therein]{Trigo2012,Fridriksson2015}. GX 13+1 is likely an X-ray dipping source. Dipping sources are binary systems with shallow X-ray eclipses, so called dippers. These eclipses are likely caused by the periodic obscuration of the central X-ray source by a structure called bulge (or hot spot), or by outflows in the outer disk \citep{Lewin1995, pintore2014}. The bulge is created by the collision between the accretion flow coming from the secondary star with the outer edge of the accretion disk, and it contains optically thick material. Therefore, the observation of dippers is associated with LMXB sources with very high inclination. The strong energy-dependent obscuration in GX 13+1, the absence of eclipses, and the spectral type of the secondary star indicate an inclination of 60-80$^{\circ}$ \citep{Trigo2012}. 

The source GX 13+1 is known as one of the two NS LMXBs showing winds \citep{Trigo2012, Ueda2004}; the other source is IGR J17480-2446 \citep{Miller2011}. A thermal driving mechanism of the wind is expected \citep[see discussion in ][and references therein]{Trigo2012}.

\subsection*{Spectral characteristics of GX 13+1}

        The continuum emission of GX 13+1 has previously been modeled with a multicolor blackbody component plus either a blackbody or a nonthermal component (such as Comptonization, power law, or cutoff power law)  without significantly worsening the quality of the fit \citep[see][and references therein]{Trigo2012, pintore2014}.

        Several narrow absorption lines, characteristics of dipping sources, are observed in the energy spectrum and are associated with a warm absorber (highly photoionized plasma) around the source, driven by outflows from the outer regions of the accretion disk \citep{pintore2014, Trigo2012}. The absorption lines indicate  bulk outflow velocities of  $\sim$ 400 km s$^{-1}$ \citep[see][and references therein]{pintore2014}. 
        
        Narrow resonant absorption lines near 7 keV were found with an ASCA observation of GX 13+1 \citep{Ueda2001}. Afterward, K$_{\alpha}$ and K$_{\beta}$ lines of He- and H-like Fe ions, the K$_{\alpha}$ line of H-like Ca (Ca XX) ion, and a deep Fe XXV absorption edge (at 8.83 keV) were observed with XMM-Newton \citep{Sidoli2002}. \citet{Trigo2012} additionally observed a Fe XXVI absorption edge (at 9.28 keV). The depth of the absorption features in the XMM-Newton 2008 observations changed significantly on timescale of a few days, and more subtle variations are seen on shorter timescales of a few hours \citep{Sidoli2002}. 
 Because the narrow absorption lines are observed throughout the orbital cycle, the absorption plasma is likely cylindrically distributed  around the compact object \citep[see][and references therein]{pintore2014}.

Another spectral feature that is superposed on the continuum and the narrow absorption lines is the broad emission line in the K-shell Fe XVIII-XXVI energy range ($\sim$ 6.4 - 7.0 keV). This emission line is the main feature we study here. It is usually modeled, as well as in all NS sources, by relativistic line models (see section \ref{sec:NSline}). 

\section{XMM Newton observations}
\label{sec:data}

        We analyzed two public XMM-Newton EPIC-pn observations of GX13+1. The first observation, Obs. ID 0505480101 (hereafter called Obs.~1), has 13.8 ks EPIC-pn exposure time, starting on 2008 March 09  at 18:24:01 and ending at 22:48:30 UTC. The second observation,  Obs. ID 0505480201 (hereafter called Obs.~2) has 13.7 ks EPIC-pn exposure time, starting on 2008 March 11 at 23:48:02 and ending at  03:15:09 UTC of the next day (see Table \ref{tabObs.} for a log of the XMM-Newton observations used in this analysis). Both observations were taken with the camera operating in timing mode. 

All the five public XMM-Newton EPIC-pn observations analyzed by \citet{Trigo2012} were previously considered for this study. However, a pre-analysis of the spectra led to the selection of the two observations used in this paper. Because all observations are affected by pile-up, we extracted two spectra for each observation, A and B, respectively. A different RAWX range was excised for each during the data processing for pile-up correction: one column was excised for A spectra from the bright central region, and either two or three RAWX columns were excised for B spectra. We fit the A and B spectra with the following total model: \textsc{tbabs*edge$_1$*edge$_2$*(diskbb+bbodyrad+gaussians)}, which corresponds to \textit{Model 1} in \citet{Trigo2012} (\textit{Model 1}; hereafter). The Gaussian components in this case correspond to negative Gaussians. We used this total model to check  for the iron emission line because the absorption lines can mimic the emission line. To select the observations, we used the following criterion: observations in which the line barely appears, that is, with a residual fluctuation < $2.5-3.0~\sigma$, when one or two columns are excised, were discarded. It is important to state that when the pile-up correction was performed excising five columns, as performed by \citet{Trigo2012}, the emission line disappears in all observations.  Therefore, we extracted two columns in the selected two observations (see section \ref{sec:datared}) to find a balance between pile-up correction and the presence of the line. The continuum is not strongly affected by the difference in the RAWX extraction in these two observations (i.e., by excising two or five columns).

The pile-up correction makes the spectra softer. Consequently, if the line is weak in the observation, it decreases the evidence of its presence. In addition, pile-up correction can deteriorate the statistics of the counts in the hard part of the spectrum, which contributes to a poorer constraining of the continuum model. This in turn affects the emission line presence perception and its shape definition.

\begin{table*}
\caption{XMM-Newton observations of GX 13+1 - basic information}
\label{tabObs.}
\center
\begin{tabular}{lllll}
\toprule
Obs.        & observation ID & Exposure Start  Time (UTC)                  & Exposure  End Time (UTC)                    & Exp.\tablefootmark{1}\\
                &                          &  year~ month day hh:mm:ss  &  year~ month day hh:mm:ss  & (ks)\\
\midrule
\vspace{1mm}
Obs.~1              & 0505480101     &  2008 March 09 ~~18:24:01  &  2008 March 09~~ 22:48:30   & 13.8\\
Obs.~2              & 0505480201     &  2008 March 11  ~~23:48:02 &  2008 March 12~~ 03:15:09   & 10.3\\
\bottomrule 
\end{tabular}
        \tablefoot{
          \tablefoottext{1}{ EPIC-pn exposure.}
        }
\end{table*}

\subsection{Data reduction}
\label{sec:datared}

All light curves and spectra were extracted through the Science Analysis Software (SAS) version 14.0.0. Following the recommendation for the pn camera on the timing mode of observation, we selected events in the 0.6-10 keV energy range to avoid the increased noise, and only single and double events were taken into account for the spectrum extraction (pattern in [1:4]). In order to provide the most conservative screening criteria, we used the FLAG=0 in the selection expression
for the standard filters, which excludes border pixels (columns with offset), for which the pattern type and the total energy is known with significantly lower precision (see XMM-SOC CAL-TN-0018, and The XMM-Newton ABC Guide).

The calibration of the energy scale in EPIC-pn in timing mode requires a complex chain of corrections, which are important to yield a more accurate energy
reconstruction in the iron line (6~-~7 keV) energy range (see XMM-SOC-CAL-TN-0083). In order to achieve the best calibration, we applied a) the X-ray loading (XRL)
correction through the parameters runepreject=yes and withxrlcorrection=yes; b) the special gain correction, which is default in SASv14.0 (and can be applied through the parameter withgaintiming=yes); and c) the energy scale rate-dependent PHA (RDPHA) correction, which was introduced in SASv13.0, and taken as default in SASv14.0 for Epic-pn timing mode (XMM-SOC-CAL-TN-0018). This correction is expected to yield a better energy scale accuracy than the rate-dependent charge transfer inefficiency (RDCTI) correction for the EPIC-pn timing mode, respectively. \citet{pintore2014} tested both rate dependence corrections in GX 13+1 in EPIC-pn timing mode. They showed that the RDPHA calibration provides more consistent centroid line energies and more physical fit parameters (such as the emission Gaussian line centroid energy and source inclination parameters obtained through the diskline model best fit). Using this correction, they found the energy of the iron mission line at 6.6 keV, and stated that it was consistent with the energies found by \citet{Trigo2012} and \citet{Dai2014}. In the RDPHA correction, the energy scale is calibrated by fitting the peaks in derivative PHA spectra corresponding to the Si ($\sim$1.7 keV) and Au ($\sim$ 2.3keV) edges of the instrumental responses, where the gradient of the effective area is largest (XMM-SOC-CAL-TN-0306).

We extracted source+background spectrum and light curves of both observations (Obs.~1 and 2) from 10 < RAWX < 60 and the background from 3 < RAWX < 9.
The SAS task \textit{epatplot} was used as a diagnostic tool for pile-up in the pn-camera. It showed that both observations are affected by pile-up. In order to mitigate its effect, we excised the brightest central columns RAWX 38-39 during the spectral extraction. 

The ancillary and response matrices were generated trough the SAS task \textit{arfgen} (in the appropriated way to account for the area excised due to pile-up) and \textit{rmfgen}. The EPIC-pn spectra were rebinned in order to have at least 25 counts in each background-subtracted channel and in order to avoid oversampling the intrinsic energy resolution by a factor larger than 3.

The light curves were produced trough the SAS task \textit{epiclccorr}, which corrects the light curve for various effects affecting the detection efficiency (such as vignetting, bad pixels, PSF variation and quantum efficiency), as well as for variations affecting the stability of the detection within the exposure (such as dead time and GTIs). Because all these effects can affect source and background light curves in a different manner, the background subtraction was made accordingly through this SAS task.

\subsection{Light curves}
\label{sec:LCs}

\begin{figure*}
\center
   \subfigure{\includegraphics[scale=0.32]{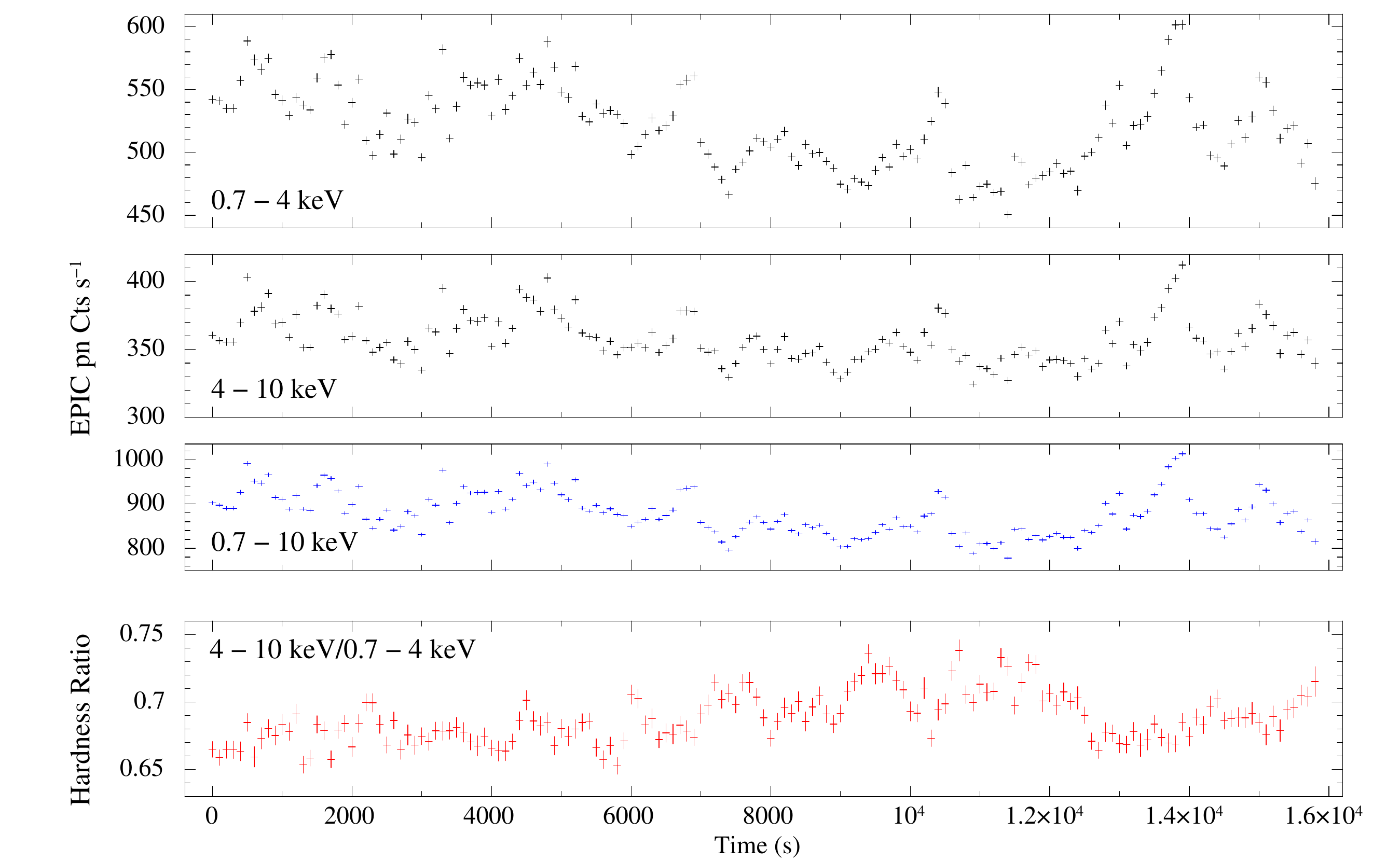}} \label{fig:LC101} 
   \qquad
   \subfigure{\includegraphics[scale=0.32]{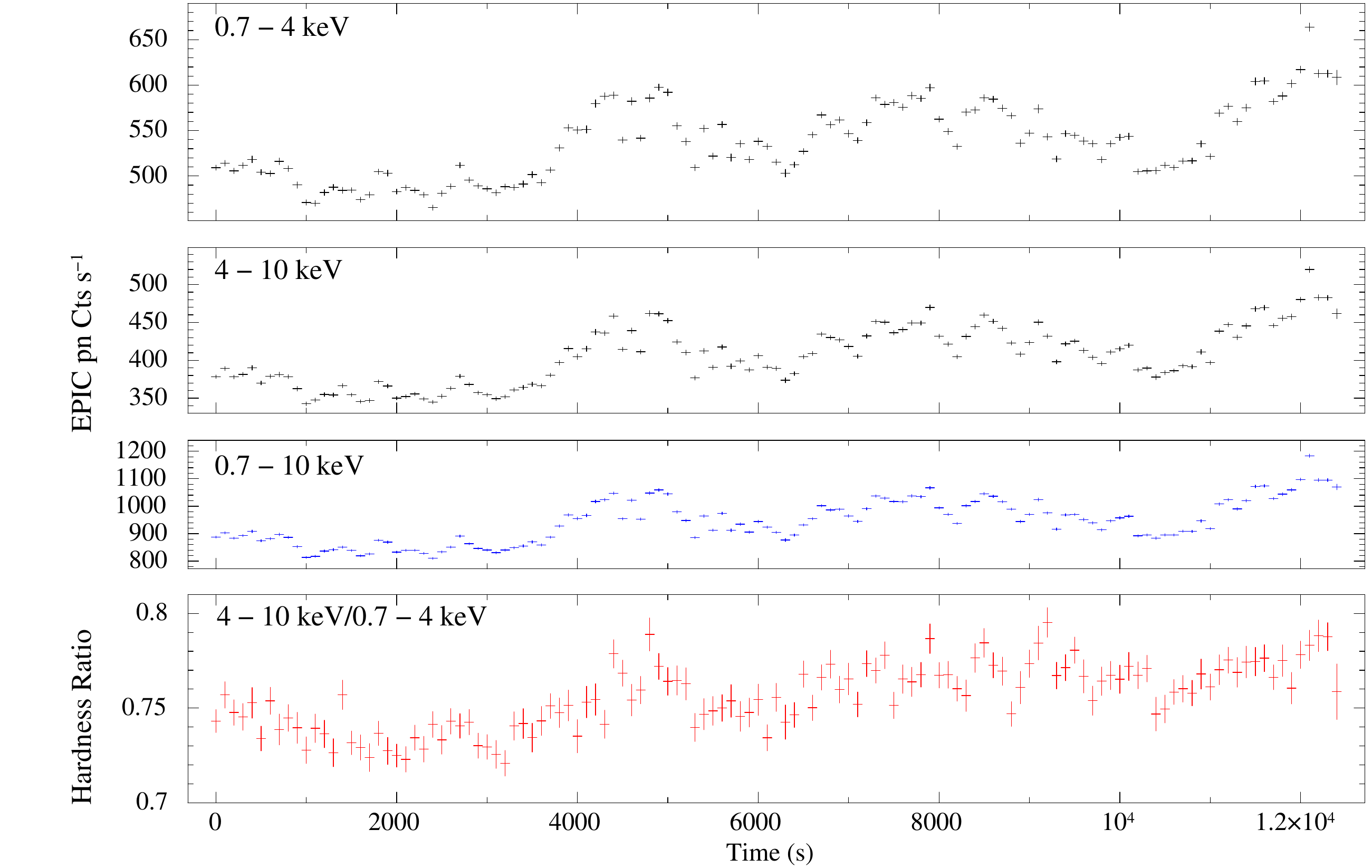}} \label{fig:LC201}
   \caption{GX 13+1 XMM Newton EPIC pn light curves and hardness ratio (with a bin time of 100.0 s) of Obs.~1 (\textit{left panel group}) and  Obs.~2  (\textit{right panel group}). In both panel groups, the upper panel shows the light curve in the  0.7 to 4.0 keV energy range, the second panel shows the light curve in the  4.0 to 10.0 keV energy range, the third panel (in \textit{blue}) shows the light curve in the total energy range, from 0.7 to 10.0 keV, and the lower panel (in \textit{red}) shows the hardness ratio between the photon counts in the 4.0 to 10.0 keV energy range by those in the 0.7 to 4.0 keV energy range.}
\label{fig:LC}  
\end{figure*}

Figure \ref{fig:LC} shows the EPIC-pn light curves extracted from Obs.~1 and 2 of GX 13+1. We present the light curves in the 0.7 - 10.0 keV,  0.7 - 4.0 keV, and 4.0 -10.0 keV energy ranges,  with a bin size of 100.0 s, and also the hardness ratio between the photon counts in the last two energy bands. Both light curves show high count rate variability.

Although we see the same shape of the light curves as in \citet{Trigo2012}, who analyzed the same EPIC-pn observations of GX13+1, we obtained a slightly higher count rate. We tried to find the origin of the discrepancy on the count rate (of $\sim$50 count/s). In order to do this, we first extracted the light curve in the same way as \citet{Trigo2012}, that is, considering the same extraction region (private communication). We obtained the same count rate as before, which means that the difference is not due to the different extraction region used. We then extracted the event list in the same way as \citet{Trigo2012}, that is, we applied the RDCTI calibration correction, and extracted the light curve without applying background subtraction, to see whether somewhat unexpectedly, the origin could be in these points. As expected, we obtained the same count rate as before and thus failed to find the origin of the discrepancy. The only possibility we see lies in the different SAS versions that were used in each analysis. We used SAS version 14.0.0 and \citet{Trigo2012} used SAS version 10.0.0. Consequently, we applied the XRL correction inserted in SAS 13.5/14.0.0. \citet{Trigo2012} calculated the ``residual'' of the offset map by subtracting the offset map of a nearby observation taken with closed filter from the offset map, and applied the XRL correction excising the inner region, which they found to be affected by XRL (which is the same region as was excised for pile-up correction). The differences in the XRL correction procedures might have contributed for the discrepancy in the count rate. Furthermore, the RDCTI calibration correction has been changed since the release of SASv10.0.0; the RDCTI correction applied by us (for the comparison) is not the same as was applied by \citet{Trigo2012}.

In Obs.~1 we observed a maximum fractional variation in the hardness ratio of $\sim$ 7.3\% with a mean value of 0.688(1). A similar variation was identified in Obs.~2, where we observed a maximum fractional variation of $\sim$ 5.2\% with a mean value of 0.76(7). We confirm what was found by \cite{Trigo2012}: the count rate variability is not associated with significant change in the hardness ratio. Because we did not find any significant change in the hardness ratio, we used the total EPIC-pn exposure time in the spectral extraction for both observations.

\subsection{Spectral variability}
\label{sec:spec}

Although the variation in hardness ratio is not significant, \citet{Trigo2012} pointed out that Obs.~1 shows a significant spectral change when the spectrum is divided into two intervals (``high and low''). However, we did not obtain this result. We checked the possible spectral variation by dividing the spectra into two parts: the first part using the good time interval (GTI) for times shorter than $\sim$ 7000 s or longer than $\sim 1.25 \times 10^4$ s, and the second part was extracted using the GTI between these two values. The spectra do not show any spectral variability.

Conservatively, we also extracted two spectra from Obs.~2: the first using the GTI in which the detected count rate was lower than 900 count $s^{-1}$ , and the second using a count rate higher than this amount. We did not find spectral variation in either observation (only the normalization parameter was different), therefore we present the spectra using the total EPIC-pn exposure time.

\section{Data analysis}
\label{sec:datanalysis}

 In our spectral analysis we used the 2.5-10.0 keV energy range. We excluded the soft energy range in which weak absorption features at $\sim$ 1.88 keV and $\sim$ 2.3 keV were observed in previous analyses and were referred to as being most likely caused by residual calibration uncertainties. We observed the feature at $\sim$ 1.88 keV in both observations (which could be Si XIII, or systematic residuals at the Si edge owing to a deficient calibration), but we did not observe the feature at $\sim$ 2.2-2.3 keV (which could be residuals of calibration around the instrumental edge of Au-M at 2.3 keV) \citep{Trigo2012,pintore2014}. The absence of the weak absorption line at $\sim$2.3 keV in our analysis could be due to improvement in the calibration in SAS version 14.0.0. Obs. 1 and 2 in this paper correspond to observations 4 and 6 in \citet{Trigo2012}.

We fit the spectral continuum with a blackbody component (\textsc{bbodyrad} in XSPEC) to model the thermal emission from the neutron star surface, plus a Comptonization component (\textsc{compTT} in XSPEC), which represents the emission attributed to a corona. The continuum of GX~13+1 is modified by significant total photon absorption by the material present in the line of sight of the observer. We used only one component (\textsc{tbabs} in XSPEC) to indicate both the Galactic absorption due to neutral hydrogen column and the variable absorbing material close to the source. The \textsc{tbabs} parameter NH was therefore let free in all fits.

\begin{figure*}
\centering
   \subfigure{\includegraphics[width=8.8cm]{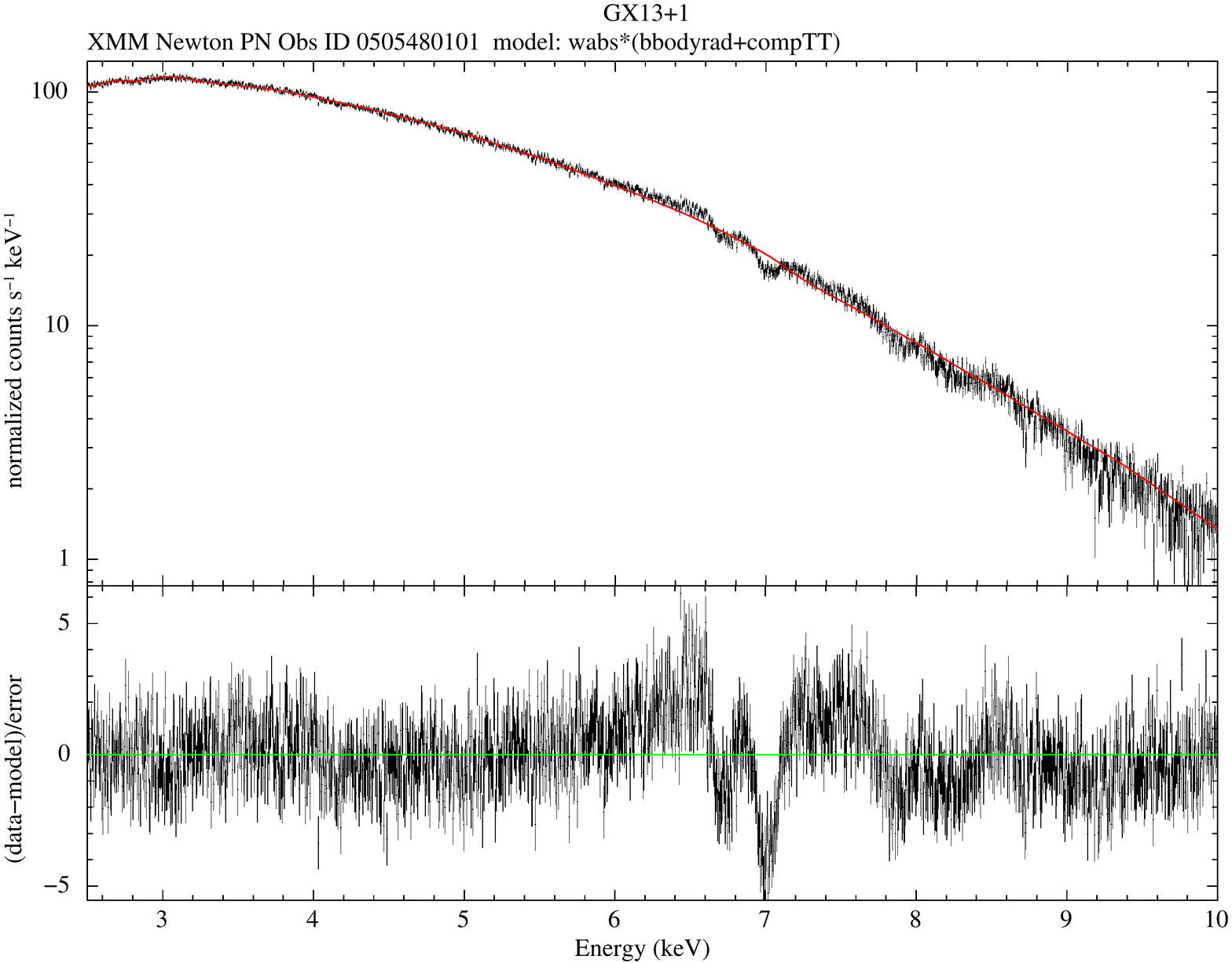}} 
   \qquad
   \subfigure{\includegraphics[width=8.8cm]{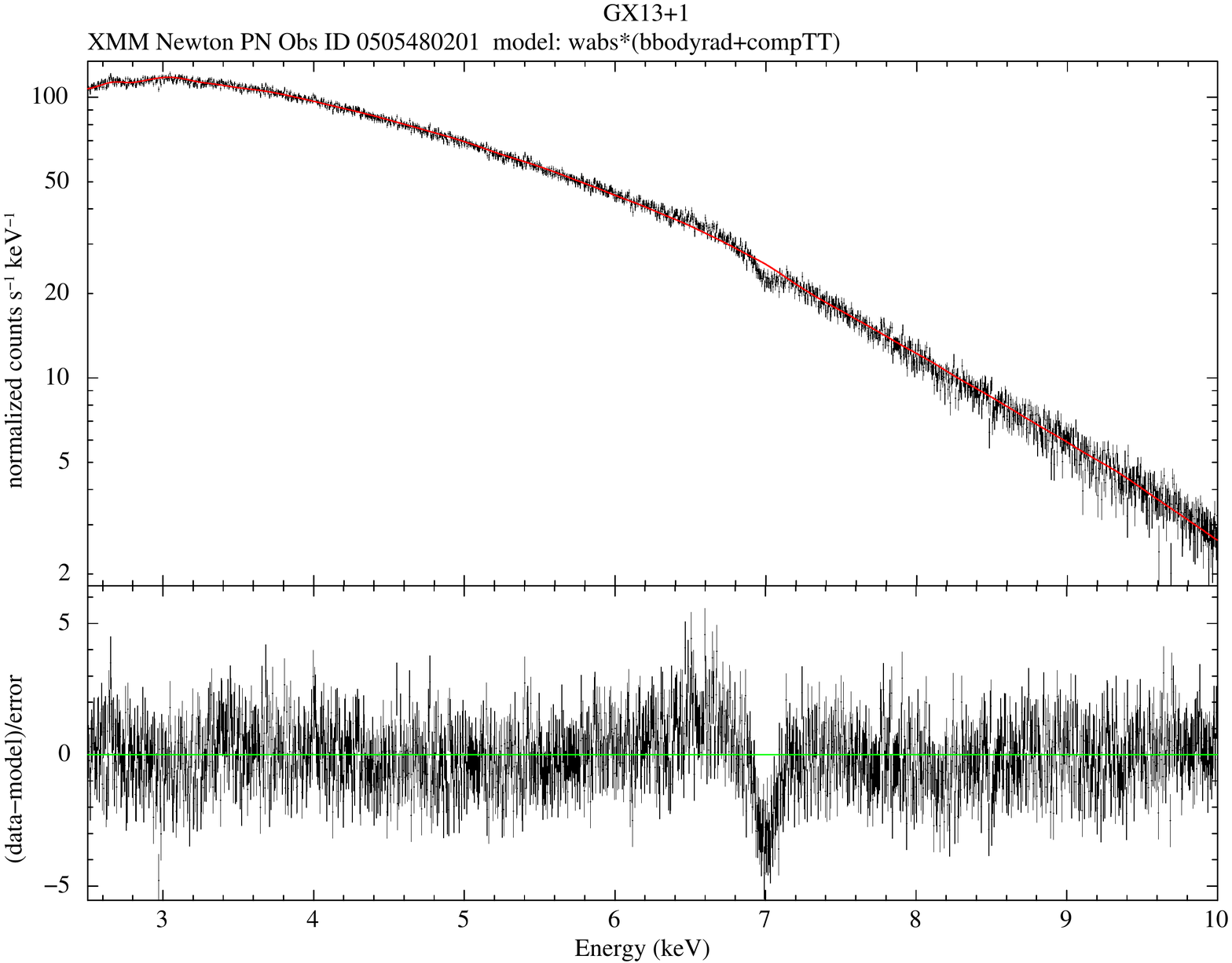}} 
   \caption{XMM Newton EPIC pn spectra of Obs.~1 (\textit{left panel}) and Obs.~2 (\textit{right panel}) of GX13+1 in the 2.5-10 keV energy range: the upper panel shows the data (\textit{in black}) and the best-fit model \textsc{tbabs*(bbodyrad+compTT)} (\textit{solid red line}), the lower panel shows the residual of the data vs. continuum model, where the emission and the absorption features are clearly visible.}
 \label{fig:emline}  
\end{figure*}

Both observations show a broad and red-skewed emission line in the Fe K$_{\alpha}$ energy band (see Figure \ref{fig:emline}). In previous analyses \citep{pintore2014, Trigo2012}, these features were fit either using a simple broad Gaussian or the relativistic \textsc{diskline} model. In this work, we fit the broad red-skewed line profile using both the nonrelativistic \textsc{windline} and the \textsc{diskline} model, and we compared the results obtained in each fit. The spectral analyses were performed using the XSPEC astrophysical spectral package version 12.09.0j \citep{Arnaud1996}. 

In the same energy band in which the Fe emission line is apparent, we observed three narrow absorption lines in Obs.~1 and one in Obs.~2. These features are characteristics of dipping sources and were modeled with absorbed Gaussians, the \textsc{lgabs} model, which differs from the \textsc{gabs} model by the fact that it is not an exponentially multiplicative model (suitable for  describing cyclotron resonance features), but a multiplicative model of the form $\propto (1-\text{Gaussian})$ \citep{Soong1990}. Because \textsc{lgabs} is a multiplicative model, it is not possible to obtain the EW of these lines through XSPEC. The addition of edges at 8.83 keV (Fe XXV) and 9.28 keV (Fe XXVI) did not improve the quality of the fit significantly. The best-fit parameters are shown in Table \ref{tab:parameters}.

In Obs.~1, the three narrow absorption lines are identified as coming from highly ionized iron: K$_{\alpha}$ Fe He-like line (Fe XXV), K$_{\alpha}$ H-like line (Fe XXVI), and K$_{\beta}$ Fe He-like line. In Obs.~2, only the K$_{\alpha}$ Fe He-like absorption line is present. The addition of a narrow absorbed K$_{\beta}$ Fe H-like line did not significantly improve the fit.

\subsection*{Fitting the broad iron emission line}

When we fit the Fe emission line with a simple Gaussian in Obs.~1, we obtained the line centroid energy $E$ equal to $6.872^{+0.030}_{-0.021}$ keV, $\sigma$ equal to $0.59^{+0.03}_{-0.05}$ keV, and the equivalent width (EW) equal to $226^{+142}_{-222}$ eV; in Obs.~2, we obtained the line centroid energy equal to $6.58^{+0.05}_{-0.05}$ keV, $\sigma$ equal to $0.64^{+0.04}_{-0.04}$ keV, and EW equal to $118^{+60}_{-87}$ eV. The centroid energies are different between the two observations. The EW of the lines are poorly constrained in both fits, and the broadening of the lines agrees within the errors between the observations at 90\% confidence level. This constancy of the line width is in agreement with what was reported by \citet{Cackett2013}.

Comparing our results with \citet{Trigo2012}, we obtained a higher Gaussian centroid energy in Obs.~1 (see Table \ref{tab:compDT12}).  However, when we fit the spectra of Obs.~1 and 2 with the same total model used by \citet{Trigo2012} (see Table \ref{tab:compDT12}, rows 1 and 3), we obtained that all parameters, except for the $\sigma$ in Obs. 1, agree at 90\% confidence level with the values found by \citet{Trigo2012}. The $\sigma$ in Obs. 1 appears slightly narrower in our analysis.

Because \citet{Trigo2012} applied RDCTI correction in their analysis, the RDPHA correction applied by us could have led to higher centroid energies, larger broadness, and higher intensity of the emission line \citep[see][]{pintore2014}. However, we did not observe these differences in the line when we fit the data with the same total \textit{Model 1} used by \citet{Trigo2012} (see Table \ref{tab:compDT12}, rows 1 and 3). Therefore, the different centroid energy found by us in Obs.~1 (see Table \ref{tab:compDT12}, row 4, column 3) is likely due to the differences in the total model used to fit the continuum and the absorbed narrow lines. However, in Obs.~2 we did not observe a shift in the line centroid energy with respect to the fit performed by \citet{Trigo2012}.

\begin{table*}
\caption{Comparison of the best-fit parameters of the iron emission line
with the previous analysis performed by \protect\citet{Trigo2012}.}
\centering
\label{tab:compDT12}
\small
\begin{tabular}{lccccccccc}
 & & \multicolumn{3}{c}{Obs.~1\tablefootmark{(a)}} & & \multicolumn{3}{c}{Obs.~2\tablefootmark{(b)}}&\\
 & & \multicolumn{3}{c}{Emission line parameters} & & \multicolumn{3}{c}{Emission line parameters}&\\
\cline{3-5} \cline{7-9}\\
Line model &Total model &E (keV) & $\sigma$ (keV) & EW (eV) & & E (keV) & $\sigma$ (keV) & EW (eV)&Reference\\ 
\toprule
\vspace{3mm} 
 Gaussian & Model 1\tablefootmark{(c)}&$6.53^{+0.11}_{-0.09}$ &$0.72^{+0.14}_{-0.16}$ &$185^{+62}_{-62}$ & &$6.63^{+0.13}_{-0.13}$ &$0.74^{+0.19}_{-0.19}$ &$108^{+27}_{-27}$ &\citep[][Table 3]{Trigo2012}\\
\vspace{3mm} 
Gaussian & Model 2\tablefootmark{(d)} &$6.56^{+0.10}_{-0.07}$ &$0.88^{+0.12^{(h)}}_{-0.07}$ &$299^{+77}_{-77}$ & &$6.71^{+0.12}_{-0.16}$ &$0.77^{+0.23^{(h)}}_{-0.15}$ &$91^{+48}_{-23}$ &\citep[][Table 4]{Trigo2012}\\
\vspace{3mm} 
Gaussian &Model 1\tablefootmark{(c)} &$6.67^{+0.06}_{-0.03}$ &$0.50^{+0.03}_{-0.10}$ &$128^{+18}_{-18}$ & &$6.57^{+0.06}_{-0.07}$ &$0.50^{+0.05}_{-0.07}$ &$78^{+18}_{-17}$ &The present paper\\
\vspace{3mm} 
Gaussian &Model 3\tablefootmark{(e)}&$6.872^{+0.030}_{-0.021}$ &$0.59^{+0.03}_{-0.05}$ &$226^{+142}_{-222}$ & &$6.58^{+0.05}_{-0.05}$ &$0.64^{+0.04}_{-0.04}$ &$118^{+60}_{-87}$ &The present paper\\
\vspace{3mm} 
Diskline    &Model 4\tablefootmark{(f)}&$6.674_{-0.002}^{+0.040}$ &$-$ &$256_{-73}^{+128}$ &$$ &$6.26_{-0.04}^{+0.05}$ &$-$ &$119_{-25}^{+21}$ &The present paper\\
\vspace{3mm} 
Windline   &Model 5\tablefootmark{(g)}&$[6.6]$ &$-$ &$195_{-47}^{+26}$ &$$ &$[6.6]$ &$-$ &$130_{-24}^{+25}$ &The present paper\\
\bottomrule 
\end{tabular}
        \tablefoot{Spectral uncertainties are given at at 90\% ($\Delta\chi^2 = 2.71$) confidence level for one derived parameter. Fixed parameters are shown in brackets.\\
                \tablefoottext{a}{XMM-Newton/PN Obs.~1 (Obs. ID 0505480101), which corresponds to Obs. 4 in \protect\citet{Trigo2012}.}\\
                \tablefoottext{b}{XMM-Newton/PN Obs.~2 (Obs. ID 0505480201), which corresponds to Obs. 6 in \protect\citet{Trigo2012}.}\\
                \tablefoottext{c}{XSPEC Model: \textsc{tbabs*edge$_1$*edge$_2$*(diskbb+bbodyrad+gau$_1$+gau$_2$+gau$_3$+gau$_4$+gau$_5$)}; where \textsc{gau$_1$} corresponds to a gaussian model, and \textsc{gau$_{(2,...,5)}$} correspond to negative gaussians.}\\
                \tablefoottext{d}{XSPEC Model: \textsc{tbabs*cabs*warmabs*(diskbb+bbrad+gau$_1$)}.}\\
                 \tablefoottext{e}{XSPEC Model: \textsc{tbabs*(bbodyrad+compTT+gaussian)*$lgabs_1$*$lgabs_2$*$lgabs_3$}.}\\     
                 \tablefoottext{f}{XSPEC Model: \textsc{tbabs*(bbodyrad+compTT+diskline)*$lgabs_1$*$lgabs_2$*$lgabs_3$}.}\\     
                 \tablefoottext{g}{XSPEC Model: \textsc{tbabs*(bbodyrad+compTT+windline)*$lgabs_1$*$lgabs_2$*$lgabs_3$}.}\\
                 \tablefoottext{h}{Parameter is pegged.}
		 }
\end{table*}

\subsection{\textsc{windline} fit}
\label{sec:windlinefit}

When we fit the \textsc{windline} model to the residual excess in the iron line energy range, we obtained the best fit-parameters shown in Table \ref{tab:parameters}; columns 4 and 5 show the best-fit parameters for Obs. 1 and Obs. 2, respectively. All fits were performed in order to leave the maximum number of free parameters.

\begin{figure*}
\center
    \subfigure{\includegraphics[width=8.8cm]{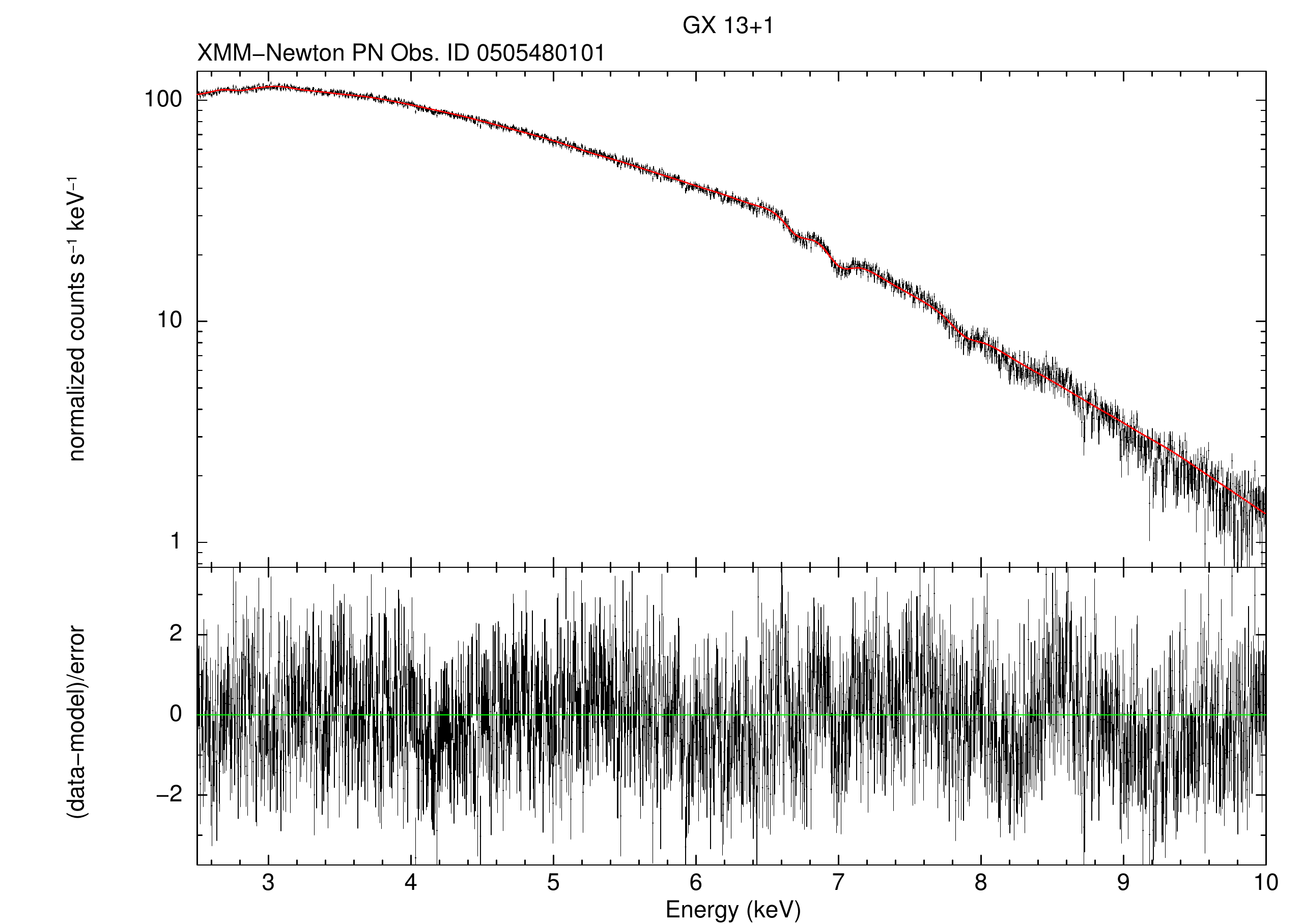}}\label{fig:SP101}
    \qquad
    \subfigure{\includegraphics[width=8.8cm]{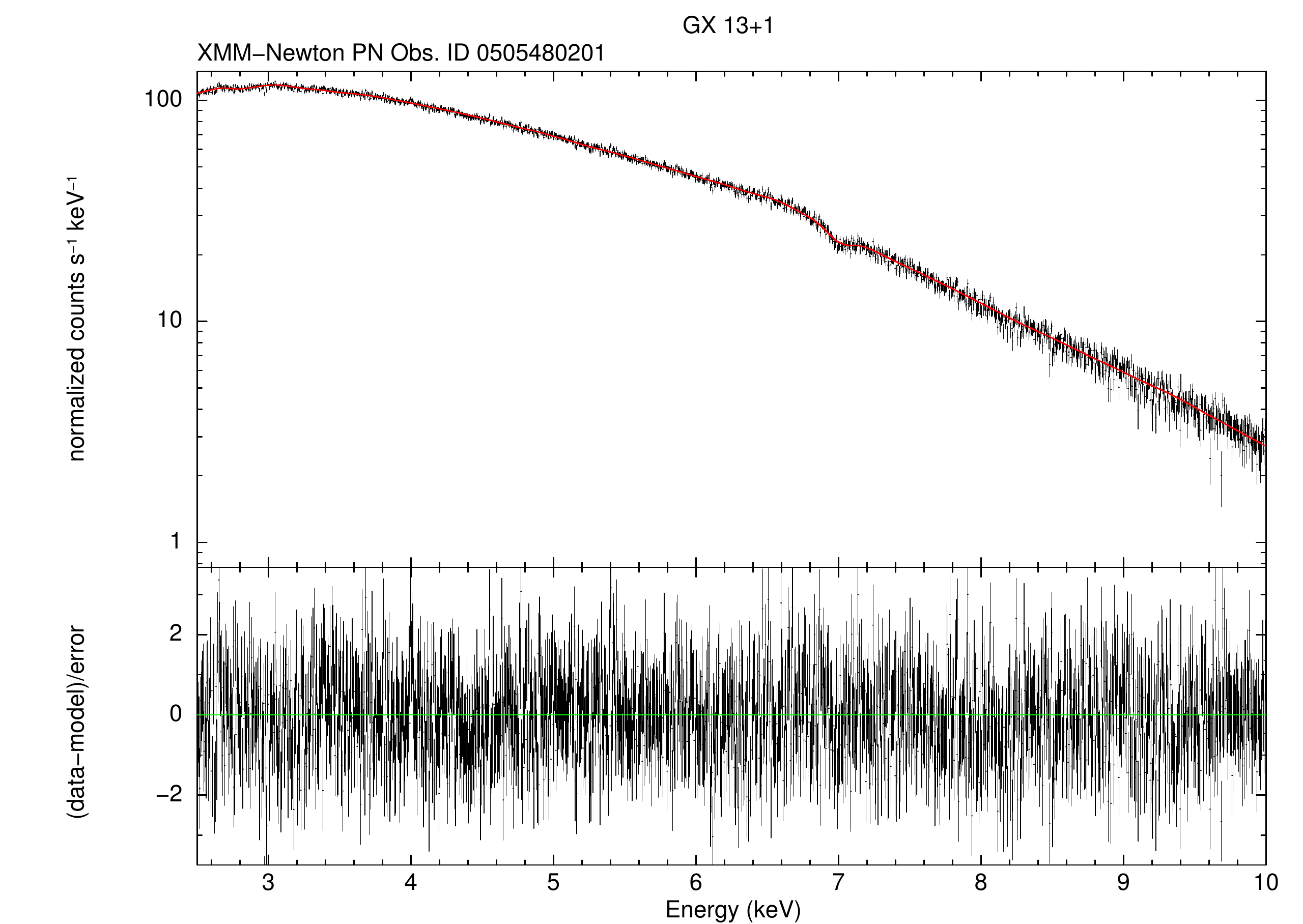}}\label{fig:SP101uf} 
   \caption{GX13+1 XMM Newton EPIC pn spectra. \textit{Left panel}: Spectrum  of Obs.~1 in the 2.5-10 keV energy range, the data (\textit{in black}), and the total model \textsc{tbabs*(bbodyrad+compTT+windline(6.6 keV))*lgabs*lgabs*lgabs} (\textit{solid red line}). \textit{Right panel}: Spectrum of Obs.~2 in the 2.5-10 keV energy range, the data (\textit{in black}), and the best-fit model \textsc{tbabs*(bbodyrad+compTT+windline(6.6 keV))*lgabs} (\textit{solid red line}). The lower panels, in the \textit{left} and  \textit{right} figures, show the residuals of the data vs. model}
 \label{fig:spec101}  
\end{figure*}

Figure~\ref{fig:spec101} shows the best spectral fit in the 2.5 to 10 keV energy range for Obs.~1 (\textit{left})  and Obs.~2 (\textit{right }) when the \textsc{windline} model was used to fit the residual excess in the iron line energy range. Figure \ref{fig:spec201} shows the unfolded spectrum in the energy range in which both the broad emission line and the narrow absorption features appear simultaneously.

The best fit was found in both observations with a 6.6 keV emission line, which corresponds to a $K_{\alpha}$ transition of the Fe XXI-XXIII ions. This line appears with an EW equal to $195_{-47}^{+26}$ in Obs. 1, and $130_{-24}^{+25}$ in Obs. 2 (see Table \ref{tab:compDT12}, \ref{tab:parameters}). We obtained the highest value of the hydrogen column in Obs.~1, which is in agreement with \citet{Trigo2012}, although different absorption column models were used by them (see Table~4 therein, where observations No. 4 and 6 correspond to Obs.~1 and~2).

In Obs.~1, the temperature of the electrons kT$_{ew}$ and the velocity of the outflow were not constrained by the fit. To obtain consistent physical parameters, we therefore fixed these two parameters (kT$_{ew}$ and $\beta$) with values similar to those found by the best fit in Obs.~2 (see Table~\ref{tab:parameters}).

In  Obs.~1 and~2 we obtained $\tau_w > 1$, $\tau > \tau_{w}$, temperature of the electrons in the wind kT$_{ew}$~$\sim$~0.6 keV, and $\beta~\sim~10^{-2}$. The velocity of the outflow is equal to $2.25 \times 10^4$ km~s$^{-1}$ in Obs.~1 and ${2.01^{+0.18}_{-0.21}  \times 10^4}$ km~s$^{-1}$ in Obs.~2.

 These outflow velocities are $\text{about ten}$  times higher than the blueshifted absorption feature velocities found in previous analyses: \citet{Trigo2012} found blueshifts in GX13+1 between $\sim$ 2100 and 3700 km s$^{-1}$; \citet{Ueda2004} indicated, based on Chandra HETGS, a bulk outflow plasma velocity of $\sim$ 400 km s$^{-1}$; and \citet{Allen2016} likewise used Chandra HETG observations and found an outflow plasma velocity  $>~500$ km s$^{-1}$.
 
\subsection{\textsc{diskline} fit} 

When we fit the \textsc{diskline} model to the residual excess in the iron line energy range, we obtained the best-fit parameters shown in Table \ref{tab:parameters}. All fits were performed in order to leave the maximum number of free parameters. Column 6 shows the best-fit parameters for Obs. 1 (Fit A, hereafter); columns 7 and 8 show the best-fit parameters for Obs. 2 when the energy of the iron emission line was not constrained (Fit B, hereafter), and when the energy line was constrained to the energy range of 6.4 to 6.97 keV (Fit C, hereafter). Figure \ref{fig:DKlines} shows the unfolded spectrum (of Fits A and C) in the energy range in which both the broad emission line and the narrow absorption features appear simultaneously.

We found an inclination of the source ($i$) equal to $60_{-6}^{+4\circ}$ in Obs.~1, $75^{\circ}$ pegged at the hard limit and $61_{-5}^{+5\circ}$ in Obs.~2, in Fits B and C, respectively. All inclination values found are in the $60 - 80^{\circ}$ inclination range expected for GX 13+1 \citep{Trigo2012}. We checked the effect of the inclination on the best-fit parameters in both observations. For inclination values within the expected range the best-fit parameters do not change significantly.

The line emissivity as a function of the accretion disk radius ($r$) is an unknown function. For simplicity, the \textsc{diskline} model assumes that the line emissivity varies as $r^{B_{10}}$, and for the most part, set $B_{10}$ is equal to $-2$ \citep[see][]{Fabian1989}. This emissivity index is expected where $r$ is approximately a few outer radii ($r_c$) of the corona.
The best fit led to a power-law index $B_{10}$ of $\sim$ $-2.3$ for both observations: $B_{10}$ is equal to $-2.38^{+0.06}_{-0.06}$ in Obs. 1 and equal to $-2.25^{+0.20}_{-0.15}$ and $-2.40^{+0.12}_{-0.10}$ in Obs. 2 in Fits B and C, respectively. 

In both observations the inner disk radius ($R_{in}$) of the reflected line component is equal to $\sim$ 10 gravitational radii ($R_g$):  $R_{in}$ is equal to $9.6^{+0.9}_{-0.8}$ $R_g$ in Obs. 1 and equal to $14^{+4}_{-7}$ and $14^{+7}_{-4}$ in Obs. 2, in the Fits B and C, respectively.

The outer radius ($R_{out}$) in Obs. 1 was fixed to 1000 $R_g$, because the thawing of the parameters led to unphysical results:  either $R_{out}$ becomes smaller than $R_{in}$, or $R_{out}$ assumes the value of $\sim$ 10.000 $R_g$. On the other hand, the $R_{out}$ parameter in Obs. 2 was left free because it appears better constrained by the fit. 

We found an emission line energy of $6.674^{+0.040}_{-0.002}$ keV in Obs. 1, which corresponds to a K$_{\alpha}$ He-like Fe emission line. On the other hand, in Obs. 2 Fit B we found an anomalous emission line energy of $6.26^{+0.05}_{-0.04}$ keV, which represents a less physical energy of an iron line. The closest iron line energy is expected to be found at 6.4 keV, which corresponds to a K$_{\alpha}$ emission line from a neutral iron atom. In Fit C we therefore allowed the line energy parameter to vary only in a  physical energy range, and in this case, the line appeared pegged at the lower limit. The line appeared with an EW equal to $256_{-73}^{+128}$ eV in Obs. 1, $119_{-25}^{+21}$ eV in Obs. 2 Fit B, and not well constrained, equal to $111_{-107}^{+18}$ eV, in Obs. 2 Fit C (see Table \ref{tab:compDT12}, \ref{tab:parameters}). 

Freezing the $R_{out}$ parameter in Obs. 2 Fit B to the same value found in Obs. 1 (i.e., to 1000 $R_g$) leads to a physical emission line energy value of $\sim$ 6.6 keV, but the inclination of the source assumes the value of $\sim$ 30$^{\circ}$, which is not consistent with the dips observed in this source. In this fitting we observed an increment of the $\chi^2$-red.\\

We also determined how different line emissivity indices affect the best-fit parameters in our analysis. We list our results below.

\begin{itemize}[label={$\bullet$}]  

\item Forcing $B_{10}$ equal to zero, which is expected for $r$ < $r_c$, led to less physical values of the source inclination ($i$ > $90^{\circ}$) and an emission line energy of $\sim$ 6.8 keV in Obs. 1. In Obs. 2 (Fit B), the emission energy line shifts to a physical value of $\sim$ 6.7 keV, but all other parameters become unconstrained and assume unphysical values, that is, $R_{out}$ becomes smaller than $R_{in}$, and the source is found to be viewed face-on ($i$~$\sim$~$3^{\circ}$).  All these fits led to an increment of the $\chi^2$-red. \textit{} \\
\item Forcing $B_{10}$ equal to -3, which is expected for $r$ beyond $r_c$, where the coronal radiation intercepts the disk only obliquely, led $R_{out}$ to assume unphysical value if it was not frozen, and the emission line energy shifts to $\sim$~6.6 keV in Obs. 1. In Obs. 2 we did not observe significant changes in the best-fit parameters. All these fits led to an increment of the $\chi^2$-red.

\end{itemize}

\subsection{\textsc{windline} versus \textsc{diskline} fit}

In both observations the best fit performed with the \textsc{windline} model was found for an emission line energy at 6.6 keV. Despite the evolution of the continuum between the two observations, the iron emission line remains remarkably constant. 
In contrast, when the excess in the energy range of the Fe K emission line was fit with the \textsc{diskline} model, different line energies were found: we found a $6.674^{+0.040}_{-0.002}$ keV emission line energy in Obs.~1, a $6.26^{+0.05}_{-0.04}$ keV emission line in Obs.~2 Fit B, and a 6.4 keV emission line pegged at the lower limit in Obs.~2 Fit C, when the line energy was constrained to a physical energy range (see Table \ref{tab:parameters}). In Obs.~2, the line energy changes considerably in relation to the \textsc{windline} model, and it is low to be emitted by an iron atom in Fit B. The best fit in this case is not able to constrain the energy line parameter to a range of physical values, probably because of the large line broadening. 

\begin{figure*}
\center
   \subfigure{\includegraphics[width=8.8cm]{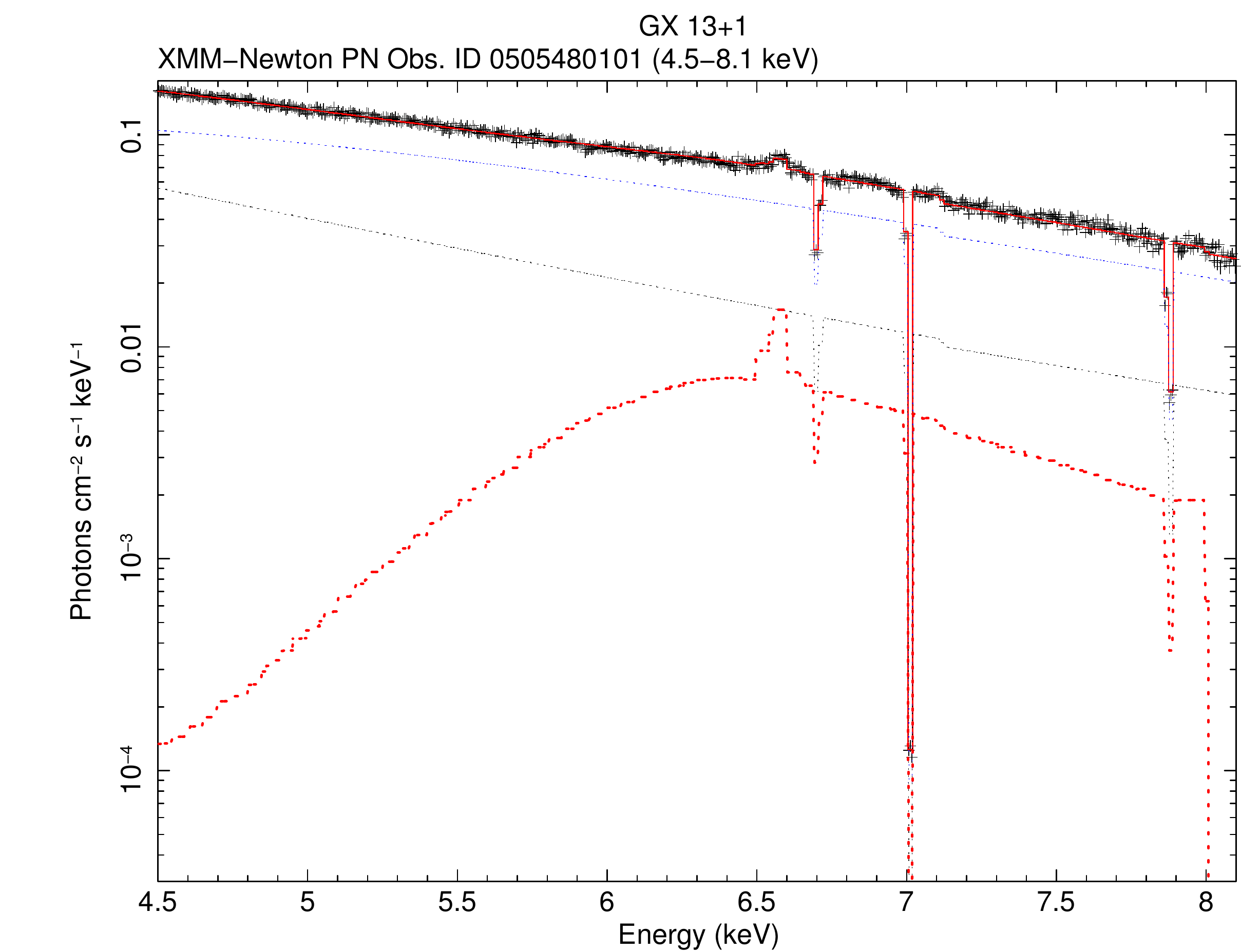}}\label{fig:SP201} 
   \qquad
   \subfigure{\includegraphics[width=8.8cm]{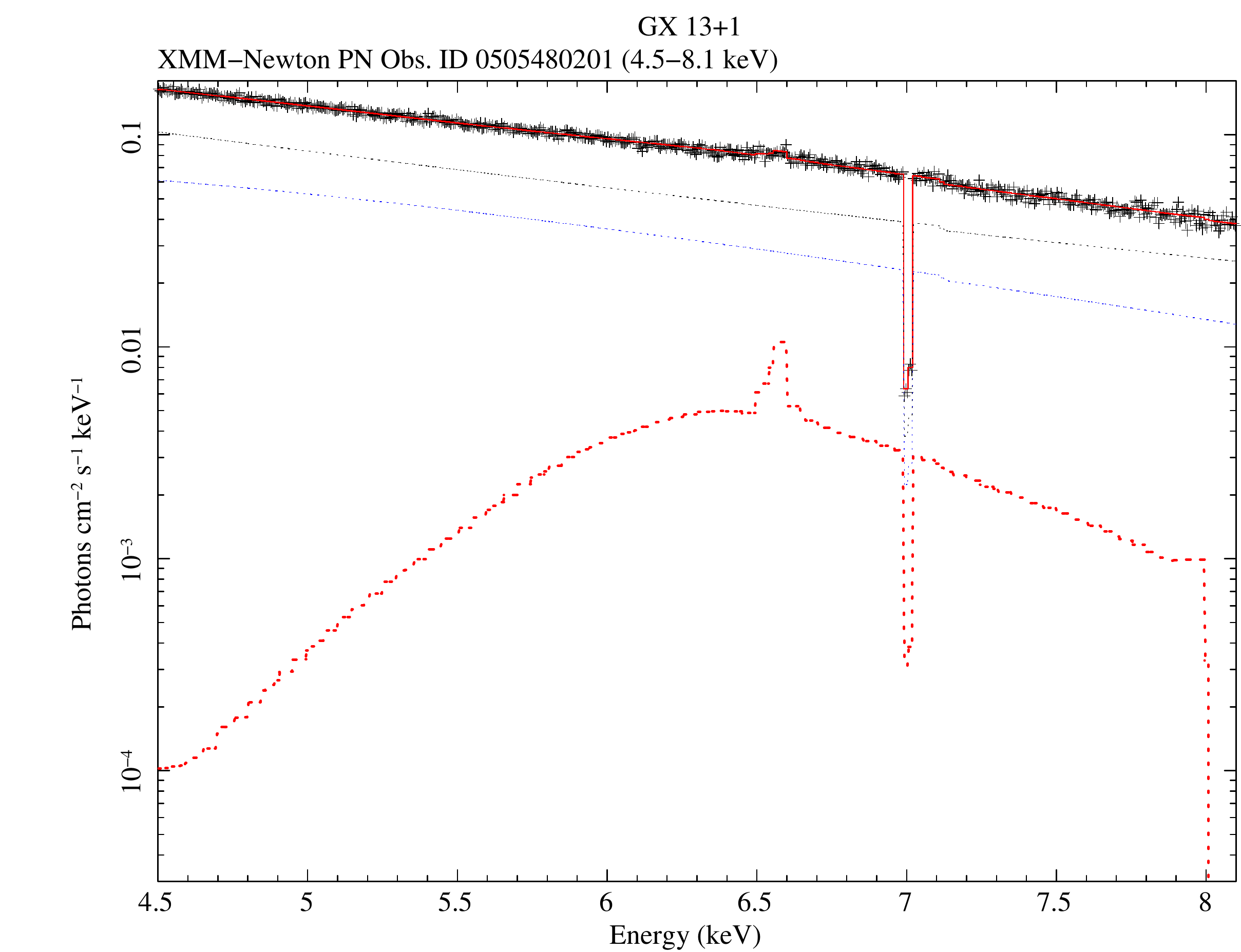}}\label{fig:SP201uf}
   \caption{GX13+1 XMM Newton EPIC pn spectra. \textit{Left panel}: Unfolded spectrum of Obs.~1  in the 4.5 to 8.1 keV energy range, the data (\textit{in black}), the total model \textsc{tbabs*(bbodyrad+compTT+windline(6.6 keV))*lgabs*lgabs*lgabs} (\textit{solid red line}), the red-skewed emission line at 6.6 keV (\textit{dashed red line}), and the continuum components (\textit{dashed black and blue lines}). The three absorbed Gaussian lines correspond to the narrow He- and H-like K$_{\alpha}$ and the He-like K$_{\beta}$ Fe absorption lines. \textit{Right panel}: Unfolded spectrum of Obs.~2 in the 4.5 to 8.1 keV energy range, the data (\textit{in black}), the total model \textsc{tbabs*(bbodyrad+compTT+windline(6.6 keV))*lgabs} (\textit{solid red line}), the red-skewed emission line at 6.6 keV (\textit{dashed red line}), and the continuum component (\textit{dashed black and blue lines}). The absorbed Gaussian line corresponds to the narrow H-like K$_{\alpha}$ Fe absorption line at 7.0 keV. }
 \label{fig:spec201}  
\end{figure*}

\begin{figure*}
\center
   \subfigure{\includegraphics[width=8.8cm]{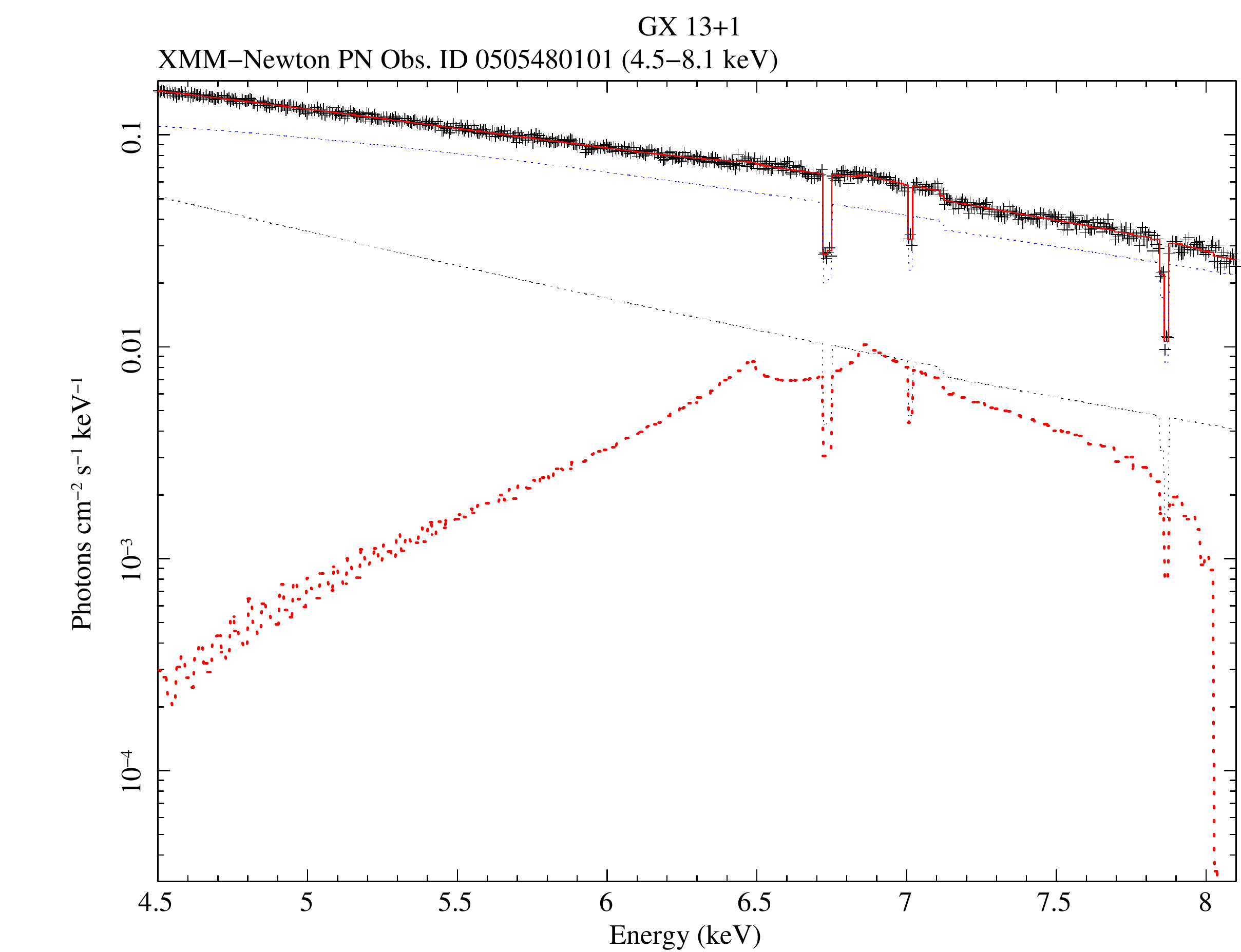}}\label{fig:DK101uf}
   \qquad
   \subfigure{\includegraphics[width=8.8cm]{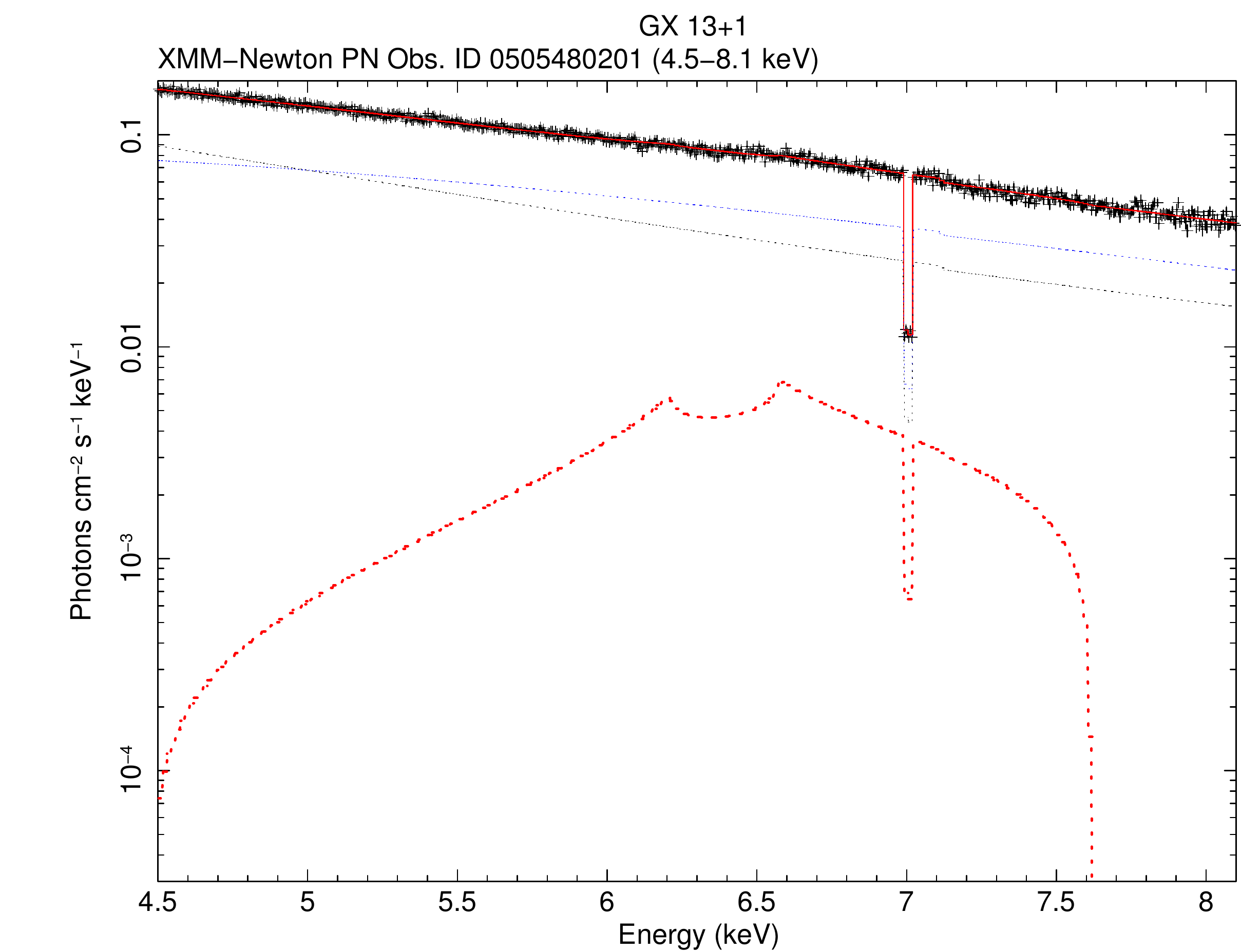}}\label{fig:DK201uf}    
   \caption{Fe emission line fit with the relativistic \textsc{diskline} model to the GX13+1 XMM Newton EPIC pn spectra. \textit{Left panel}: Unfolded spectrum for Obs.~1 in the 4.5 to 8.1 keV  energy range, the data (\textit{in black}), the total model (Fit A) \textsc{tbabs*(compTT+diskline)*lgabs*lgabs*lgabs} (\textit{solid red line}), and the Fe emission line (\textit{dashed red line}). \textit{Right panel}: Unfolded spectrum for Obs.~2 in the 4.5 to 8.1 keV energy range, the data (\textit{in black}), the total model (Fit C) \textsc{tbabs*(compTT+diskline)*lgabs} (\textit{solid red line}), and the Fe emission line (\textit{dashed red line}).}      
 \label{fig:DKlines}  
\end{figure*}

The reconstructed shape of the line also changes considerably between the two models. The line profile described by the \textsc{diskline} model (see Figure \ref{fig:DKlines}) appears double-peaked as a result of the Doppler effects on the accretion disk. The high disk inclination of the source produces a broader line in which the two peaks are very well separated. Aberration, time dilation, and blueshift by the fraction of the disk that is relevant together with the disk inclination make the blue horn brighter than the red horn \citep{Fabian1989}. In contrast to the relativistic line profiles, the \textsc{windline} line profiles (see Figure~\ref{fig:spec201}) show a broad single-peaked line, whose peak corresponds to the direct component of the line.  

All fits with the \textsc{diskline} model determined the values of the source inclination in the range expected for GX~13+1, and in Obs. 1 and Obs. 2 Fit C the inclination parameter remains constant at 90\% confidence level. The \textsc{windline} model shows no inclination effect on the line profile, the flow is spherically symmetric, with radial velocities, which means that there is no preferred angle between the flow stream and the line of sight, that is, observers from any direction see the same modified line, produced by a mean Doppler shift over all flow directions. 

We obtained that not only the energy of the emission line was affected when different line models were used. It also slightly affected the continuum and the centroid energy of the narrow He-like Fe K$_{\alpha}$ absorption line (see Table~\ref{tab:parameters} and the discussion below). 

Of the continuum parameters, in Obs.~1 only the \textsc{bbodyrad} temperature remains in agreement at the 90\% confidence level, when different line models are used. In Obs.~2, however, only the value of the hydrogen column remains in agreement at the 90\% confidence level (see Table~\ref{tab:parameters}). 

The EWs of the broad iron emission line in Obs.~1 and Obs.~2 are poorly constrained, which is partially due to the presence of the absorption lines that intercept the broad emission line, as previously noted by \citet{Trigo2012}. The width of the narrow absorption lines are not constrained by the fits either. Because of the error bars, we therefore cannot infer any change in the EW of the emission line when different line models are used in an observation. 

\citet[see Figure~3 therein]{Cackett2013}, using the results of \citet[][see Table 4 therein]{Trigo2012}, found no statistically significant correlation between the warm absorber column density and Fe emission line EW in GX 13+1. 
 They obtained that a change in the measured emission line EW with increasing N$_H$ depends on the continuum parameters. That may explain the different values of EW among Obs.~1 and~2 at the 90\% confidence level in the fits performed with the \textsc{diskline} model. This may explain the different values of EW among Obs.~1 and~2 at the 90\% confidence level in the fits performed with the \textsc{diskline} model.

\subsubsection*{Narrow and absorbed Gaussians}

When we fit the spectra with the total model used by \citet{Trigo2012}, that is, using Gaussians with negative normalizations to fit the absorbed lines, we obtained that in Obs. 1 the EW ($30 \pm 4$ eV) of the Fe He-like absorption line at $\sim$ 6.7 keV  is slightly greater  than the value found by \citet{Trigo2012} ($20 \pm 5$ eV). For all other absorbed lines in Obs.1 and for the line in Obs.2, the EWs  agree at the 90\% confidence level with \citet[][see Table 3 therein]{Trigo2012}.
The values of the centroid energies of the narrow and absorbed Gaussians found in the two observation for the \textsc{windline} and \textsc{diskline} fits are compatible with the values found by \citet{Trigo2012} and \citet[][for the RDPHA correction, although a different observation was used by them]{pintore2014} at the 90\% confidence level.

However, comparing the line energies obtained in the fits performed with the \textsc{diskline} model with the fits performed with the \textsc{windline} model, we observed that in Obs.~1 the He-like Fe K$_{\alpha}$ line appears to be blueshifted when the \textsc{diskline} model is used to fit to the broad iron emission line, whereas in the \textsc{windline} fit, it was not possible to distinguish between a blueshifted, redshifted, or not shifted (affected) line because of the error bars of this parameter. The other two narrow lines, H-like Fe K$_{\alpha}$ and He-like Fe K$_{\beta}$, showed the same ambiguity in the line shift in both observations for the \textsc{windline} and \textsc{diskline} fits.

When the narrow line indeed appeared to be shifted, the velocity of the warm absorber producing the shift in the line energy is around ten times lower than the velocity of the more central \textsc{outflow (wind)} where the iron emission line is formed in the \textsc{windline} framework. For all other cases, in which the ambiguity in the line shift is found, when any shift is present, the radial velocity of the outflow is around 100 times lower than the central outflow velocity.

\subsection{Run-test and statistical assessment of the goodness-of-fit}

Table~\ref{tab:NSruntest} shows (see columns 6, 7 and 8) that all the different line models are statistically equivalent on the basis of the 
$\chi^2$ goodness-of-fit test. For example, in Obs. 1 the F-test gives a probability of chance improvement of 16$\%$, when instead of the \textsc{windline} model the \textsc{diskline} model is used to describe the data. In Obs. 2 the F-test gives a probability of chance improvement of 48$\%$ and 41$\%$ when Fits B and C are compared with the \textsc{widnline} fit, respectively. In order to distinguish this ambiguity between the models, we used a different approach to this problem. 

Because our main goal is to determine whether the iron line profile is intrinsically asymmetric (i.e., the skewness coming from physical grounds) and to distinguish
between the relativistic and nonrelativistic cases, we need a statistical test that takes the \emph{\textup{shape}} of the feature into account. To this aim, we used the run-test, also known as the Wald-Wolfowitz test \citep{Barlow1989,Eadie1971}, which inspects the residuals of the fit. If the model perfectly describes the data, the positive ($+$) and negative ($-$) residuals are expected to be randomly distributed around zero, and the number of runs, that is, the number of sequences of consecutive
$+$ or $-$ residuals, is expected to be large \citep[see, e.g.,][for an application of the run-test in an astrophysical context]{Redman2009,Orlandini2012}. 

The run-test gives the cumulative probability of obtaining by chance the number of observed runs (run-test probability, hereafter). For a good fit, that is, for a random distribution of the residuals, the run-test probability will be high. If, on the other hand, the fit does not describe the shape of the feature, the data will not be randomly distributed with respect to the fitting model. In this case, the run-test probability will be lower, meaning that there is a residual underlying trend.

It is important to mention that the run-test and the $\chi^2$-statistic are independent. The $\chi^2$-statistic does not depend on the ordering of the bins or on the signs of the residuals in each bin, while the run-test accounts for both the ordering and the signs, providing effectively additional information on the model goodness \citep{Eadie1971}.

Table~\ref{tab:NSruntest} also shows the run-test probabilities in the 6.0 to 6.7 keV emission line energy range for the Gaussian (column 3), the \textsc{windline} (column 4), and the \textsc{diskline} (column 5) fits. For the Gaussian fit we obtained a run-test probability of $12.6\%$ in Obs.~1, and $55.2\%$ in Obs.~2. In the case of the \textsc{diskline} fit, we found a run-test probability of  $14.4\%$ in Obs.~1, $43.0\%$ in Obs.~2 when the line energy was not constrained in the fit, and $32.4\%$ when the line energy was constrained in the 6.4 - 6.97 keV energy range. When we fitted the data with the \textsc{windline} model, this probability becomes $20.0\%$ in Obs.~1 and $56.9\%$ in Obs.~2. For all line models (\textsc{gaussian}, \textsc{diskline,} and \textsc{windline}) the hypothesis that the residuals are randomly distributed is accepted for a test at $5\%$ significance level.

\begin{table*}
\caption{Best fits to the 2.5 to 10 keV EPIC pn spectra of GX13+1.}
\centering
\label{tab:parameters}
\small
\begin{tabular}{llllllll}
\hline 
\toprule
Component    & Parameter& Unit                                        & Obs.1\tablefootmark{(a)}   & Obs.2\tablefootmark{(b)} & Obs.1\tablefootmark{(a)} & Obs.2\tablefootmark{(b)} & Obs.2\tablefootmark{(b,c)}\\
\midrule
\vspace{1.5mm}  
Tbabs           & nH              & 10$^{22}$ atoms cm$^{-2}$  & $4.90_{-0.15}^{+0.15}$       & 2.9$_{-0.3}^{+0.6}$         &  $6.06_{-0.06}^{+0.40}$ & 2.48$_{-0.27}^{+0.17}$              &$2.48_{-0.39}^{+0.12}$\\  
\vspace{1.5mm}  
bbodyrad    & $kT_{bb}$   &  keV                                       &  $1.221_{-0.008}^{+0.003}$  & 1.29$_{-0.03}^{+0.03}$  &  $1.220_{-0.003}^{+0.003}$  & 1.505$_{-0.050}^{+0.023}$ &$1.503_{-0.040}^{+0.025}$\\  
\vspace{1.5mm}  
CompTT     & $kT_0$       & keV                                        &  $0.132_{-0.022}^{+0.018}$  & 0.51$_{-0.06}^{+0.03}$   &  $0.188_{-0.008}^{+0.060}$ & 0.644$_{-0.019}^{+0.025}$ &$0.65_{-0.01}^{+0.04}$\\  
\vspace{1.5mm}   
                   & $kT_e$       & keV                                        &  $2.015_{-0.009}^{+0.007}$  &  2.46$_{-0.17}^{+0.60}$  &  $2.503_{-0.003}^{+0.018}$ & 3.84$_{-0.17}^{+0.22}$       &$3.84_{-0.23}^{+0.10}$\\  
\vspace{1.5mm}   
                   & $\tau$         &                                               &   $3.80_{-0.15}^{+0.40}$       & 5.7$_{-1.0}^{+0.6}$          &  $2.54_{-0.21}^{+0.08}$  & 3.01$_{-0.20}^{+0.40}$            &$3.00_{-0.16}^{+0.40}$\\  
\vspace{1.5mm}       
Windline       & $\tau_w$    &                                             & $2.55_{-0.21}^{+0.08}$         & 2.57$_{-0.18}^{+0.15}$    &  & &\\  
\vspace{1.5mm}    
                   & $kT_{ew}$  & keV                                       & [0.7]                                        &0.6$_{-0.4}^{+0.7}$           & & &\\  
\vspace{1.5mm}    
                   & $\beta$       &                                              & [7.5$\times 10^{-2}$]              & 6.7$_{-0.7}^{+0.6}\times 10^{-2}$  &  &\\  
\vspace{1.5mm}         
Diskline      & $E_{L}$     & keV                                        &                                                &                                           &$6.674_{-0.002}^{+0.040}$   &6.26$_{-0.04}^{+0.05}$     &$6.4$\tablefootmark{(d)}  \\  
\vspace{1.5mm}
                   &  $\beta_{10}$\tablefootmark{(e)}   &          &                                                &                                            &$-2.38_{-0.06}^{+0.06}$   &-2.25$_{-0.15}^{+0.20}$        &$-2.40_{-0.10}^{+0.12}$\\  
\vspace{1.5mm} 
                   &  $R_{in}$     &  $R_G$                               &                                                &                                            &$10_{-2}^{+3}$   &14$_{-7}^{+4}$                                    &$14_{-4}^{+7}$\\  
\vspace{1.5mm} 
                   &  $R_{out}$   & $R_G$                                &                                                &                                            &$[1000]$   &413$_{-111}^{+212}$                                    &$956_{-319}^{+510}$\\  
\vspace{1.5mm}   
                  &  $i_{out}$   & deg                                        &                                                &                                            &$60_{-6}^{+4}$   &75\tablefootmark{(d)}                          &$61_{-5}^{+5}$\\  
\vspace{1.5mm}                         
EW             &                    & eV                                        &  195$_{-47}^{+26}$                &  130$_{-24}^{+25}$             &$256_{-73}^{+128}$ & $119_{-25}^{+21}$                      &$111_{-107}^{+18}$\\  
\vspace{1.5mm}                                   
Gaus$_1$  & $E_{L}$       & keV                                     &  6.701$_{-0.014}^{+0.008}$    &                                            &$6.733_{-0.013}^{+0.003}$ &                                          &\\  
\vspace{1.5mm}   
                   & $\sigma$    & keV                                      &  3.9$\times 10^{-3}$\tablefootmark{(d)}   &                            &2.8$\times 10^{-3}$\tablefootmark{(d)}  &                       &\\  
\vspace{1.5mm}        
Gaus$_2$  & $E_{L}$      & keV                                       &  7.005$_{-0.015}^{+0.015}$   & 7.005$_{-0.014}^{+0.014}$ & $7.005_{-0.015}^{+0.002}$ &7.005$_{-0.015}^{+0.014}$ & $7.004_{-0.014}^{+0.010}$\\  
\vspace{1.5mm}    
                   & $\sigma$    & keV                                       &  1.1$\times 10^{-4}$\tablefootmark{(d)} & 1.4$\times10^{-3}$\tablefootmark{(d)} & 1.1$\times 10^{-4}$\tablefootmark{(d)} & $6.7\times10^{-4}$\tablefootmark{(d)} &$1.1\times10^{-4}$\tablefootmark{(d)}\\  
\vspace{1.5mm}                        
Gaus$_3$  & $E_{L}$      & keV                                       & 7.877$_{-0.018}^{+0.015}$     &             &$7.860_{-0.015}^{+0.015}$  & &\\  
\vspace{1.5mm}    
                   & $\sigma$    & keV                                       & 2.7$\times 10^{-3}$\tablefootmark{(d)} &            &$1.6\times10^{-4}$\tablefootmark{(d)}   & &\\  
\toprule 
Fit quality   & $\chi^2$/d.o.f   &                                         & 1784.81/1481                         & 1572.35/1485 &  1690.81/1478 &1573.37/1483 &1588.06/1483\\  
\vspace{1.5mm}   
                  & $\chi^2_{red}$  &                                         & 1.21                                        & 1.06                & 1.14  &1.06 & 1.07\\                                                                               
\bottomrule 
\end{tabular}
        \tablefoot{Spectral uncertainties are given at the 90\% ($\Delta \chi^2 = 2.71$) confidence level for one derived parameter. The frozen parameters are shown in parentheses.\\
                \tablefoottext{a}{XMM-Newton/PN Obs.~1 (Obs. ID 0505480101).}\\
                \tablefoottext{b}{XMM-Newton/PN Obs.~2 (Obs. ID 0505480201).}\\
                \tablefoottext{c}{Fitting constraining the emission line in the 6.4-6.97 keV energy range.}\\
                \tablefoottext{d}{Parameter is pegged.}\\
                \tablefoottext{e}{Radial dependence of the line emissivity (power-law dependence, see \citet{Fabian1989}).}
         }
\end{table*}

\begin{table*}
\centering
\caption{Run-test on the fit residuals of the broad emission Fe K line of the NS LMXB GX~13+1.}
\label{tab:NSruntest}
\tiny
\begin{tabular}{lccccccccc}
\toprule
	Obs.  & Energy &  \multicolumn{3}{c}{Run-test\tablefootmark{(a)} (\%)} & & \multicolumn{3}{c}{$\chi^2_\text{red}$ ($\chi^2$/dof)}\\ 
\cline{3-5} \cline{7-9}   
          &  (keV) & Gaussian & windline & diskline                  &  & Gaussian & windline & diskline \\         
\midrule             
1 & 6.0--6.67  &12.6  &20.0  &14.4                                      &  & 1.16 (1714.58/1480) & 1.21 (1784.81/1481) & 1.14 (1690.81/1478)\\
	2 & 6.0--6.67  & 55.2 & 56.9 & 43.0\tablefootmark{(b)}     & & 1.06 (1578.99/1486) & 1.06 (1572.35/1485) & 1.06 (1573.37/1483)\tablefootmark{(b)}\\
	2 & 6.0--6.67  &         &         & 32.4\tablefootmark{(c)}   &  &  &  & 1.07 (1588.06/1483)\tablefootmark{(c)}\\
\bottomrule
\end{tabular}  
        \tablefoot{
        \tablefoottext{a}{Probability that n consecutive data points in the emission line region are mutually independent.}\\
         \tablefoottext{b}{Value corresponding to the fit in which the emission line energy is not constrained (see column 7 Table \ref{tab:parameters}).} \\
         \tablefoottext{c}{Value corresponding to the fit in which the emission line is constrained in the 6.4-6.97 keV energy range (see column 8 Table \ref{tab:parameters}).}
         }
\end{table*}

\section{Discussion and conclusions}
\label{sec:dis}

The fluorescent iron line is produced by hard X-ray irradiation of a cold gas. However, where and how the asymmetry of the line is created differs between the two asymmetric Fe emission line models.  

Because of the high disk inclination and the assumption that the line is created in the inner part of the accretion disk, the line profile appears to be double-peaked when it is fit with the \textsc{diskline} model. On the other hand, the line profile appears to be single-peaked when it is fit with the \textsc{windline} model because the red wing is produced by the indirect component of the line photons, which interact multiple times with the electrons in the outflow before escaping to the observer. Nonetheless, the different line model frameworks are both capable in terms of $\chi^2$-statistic to fit the broad Fe K emission line profiles. 

The Compton bump in the $\sim$ 10 - 40 keV  energy range is expected to be observed in the reflection scenario, and therefore in the relativistic line scenario \citep{Fabian2000}. However, the bump is expected in the \textsc{windline} framework as well. \citet{Laurent2007} used Monte Carlo simulations of the continuum spectrum emerging from a pure scattering wind, and their analytical description showed that when the incident spectrum is described by a hard power law ($\Gamma < 1$) and the wind has physical parameters such that $\beta \lesssim  0.3$ and $\tau$ of a few, a photon accumulation bump at energies around $\gtrsim ~10$ keV and a softening of the spectrum at higher energies are observed. Because the number of photons is conserved, the down-scattered high-energy photons, which are removed from the high-energy part of the incident spectrum, are detected at lower energy. They pointed out that as the optical depth increases, more prominent bumps are formed in the outflow. In general, the shape of the emerging continuum depends on the mean number of scatterings suffered by the photons in the outflowing plasma (which is a combined effect of the wind optical depth $\tau$ and the wind velocity), and on the incident spectrum shape.

The \textsc{windline} model allows determining the possible outflow physical parameters. For both observations, we found the

\begin{itemize}
\item optical depth of the wind ($\tau_w) > 1$;
\item temperature of the electrons in the wind (kT$_{ew})$ of ~$\sim$~0.6 keV, which is the possible outflow temperature \citep{Laming2004}; \textit{\textup{and an}} 
\item outflow velocity ($\beta$) of ~$\sim~10^{-2}c$ (see sec.~\ref{sec:windlinefit}). 
\end{itemize}

The outflow velocities determined by the \textsc{windline} model $\text{are}$ ten times higher than the velocity determined by the blueshift velocities of the absorption features found in previous analyses. The different velocities can still be explained considering that the emission and absorption lines are produced in different regions: the emission line might be produced in a more internal region of the source, at the bottom of the wind or outflow (in an inner shell) where the velocity is high; on the other hand, the resonance absorption lines are produced in a more distant region from the compact object by the interaction of photons with highly ionized species present in a cylindrical absorbing plasma around the source, driven by outflows from the outer regions of the accretion disk. Therefore, the presence of the outflow and its velocity variation along the cloud radius may explain the simultaneous observation of narrow absorption and broad emission lines in sources with high inclination, such as GX13+1. The absence of absorption lines in the internal outflow, responsible for red-skewing the line, can be explained by relatively high temperature of the outflow, which is on the order 0.3-0.6 keV, see the calculations of the temperature structure of the outflow in \citet{Laming2004}.

The wind can be launched from 3 to 10$^4$ R$_g$. In the \textsc{windline} model, the wind is considered as a spherical shell of internal radius of  100 R$_g$ and external radius of 120 R$_g$, so that the wind starts at 100 R$_g$. These values have no great influence on the emitted spectrum as long as the inner radius of the wind is far from the central compact object, where outflowing plasma is not fully ionized and strong gravitational effects are not expected.

The critical mass outflow rate \.{M} is given by $4 \pi r^2 \rho(r) m_{p} \beta c = 1.7 \beta \tau 10^{21}$ g/s = $2.6 \beta \tau 10^{-5}$ M$_{\odot}$ yr$^{-1}$ (where m$_{p}$ is the mass of the proton) \citep[see][appendix E]{TS2007}. For the fit values we found, $\tau \approx 2$ and $\beta \approx 0.01$, we find \.{M} = 3 $\times$ 10$^{19}$ g/s, which is more than ten times what was previously found \citep[see][and references therein]{Trigo2012}. However, this is consistent with the velocity difference between the inner and outer parts of the flow, outer parts where absorption lines take place. The given mass outflow rate is an upper limit because it was calculated considering that the wind is spherical throughout the system. A lower mass outflow rate may be emitted by a wind that partially covers the system. The mass outflow rate was self-consistently calculated by \citet{Laming2004} and reproduced by \citet{Laurent2007} using Monte Carlo simulations.

When the different asymmetric line models were fit to the same observation, slight differences in the continuum were observed, and the EW remained the same at the $90\%$ confidence level. The fits with the two different line models to the complex spectrum of the dipping source GX13+1 also introduced differences in the energy of the fluorescent Fe emission line and in the energy of the K$_\alpha$ Fe He-like (Fe XXV) absorption line.

The best fit with the \textsc{diskline} model in Obs.~2 gives an emission line energy equal to 6.26$^{+0.05}_{-0.04}$ keV, which is lower than the energy expected from photons coming from a neutral or ionized iron atom. It is, for example, 2.8$\sigma$ from the K$_{\alpha}$ energy line emitted by a neutral iron, at 6.4 keV. We checked if the total continuum model could affect the emission line energy in this observation, but fitting the continuum and the absorbed Gaussians as in \textit{Model 1} , that is, with \textsc{tbabs*edge$_1$*edge$_2$*}(\textsc{diskbb+bbodyrad+diskline+gaus$_1$}) model in XSPEC, did not lead to a significant change in the emission line energy (in this case, the energy of the line is equal to $6.28^{+0.04}_{-0.10}$). To obtain a physically consistent fit, we constrained the line energy parameter to the expected energy range. In this case, although the energy line appears pegged at the lower limit, the geometric parameters of the disk are found in good agreement with the fit in Obs. 1.

In the fits using the \textsc{windline} model, the outflow parameters, and consequently the properties of the inner portion of the outflow, are approximately constant in the two observations. The emission line generated in the outflow remains remarkably constant although the underlying continuum evolves. However, as in the \textsc{diskline} fit to Obs. 2, in Obs. 1 two \textsc{widnline} parameters  ($kT_{ew}$ and $\beta$) had to be frozen to obtain a physically consistent fit. The constant properties of the inner outflow may be explained by the intrinsic variability of the accretion flow.

The ambiguity in terms of $\chi^2$-statistic test between the emission line models was expected and confirmed in the spectral fits presented in this paper.  The $\chi^2$-statistic was not sufficient to lead to an unambiguous statement on the line profiles because it squares the differences between data and model and consequently looses the information about the form of the residuals in the emission line energy range. Therefore, we used the statistical run-test to take the shape of the residuals into account.
We tried to break the degeneracy between the relativistic and nonrelativistic line models in the GX 13+1 spectra because it is found in the literature that this source presents clear evidence of both disk-wind (outflow) and strong, broad, and skewed iron emission lines. However, this source has a complex spectrum, with narrow absorption iron lines in  the same energy range as the broad emission line. The absorbed iron lines complicate a distinction between the two line models because the run-test is not performed in the entire energy range of the emission line.

For the two observations analyzed in this paper, we were not able to reject the hypothesis that the residuals are randomly distributed at the $5\%$ significance level, and we cannot rule out one of the line models. 
We obtained a higher cumulative probability of observing by chance the number of runs around the fitting line for the \textsc{windline} spectral fit. However, when we modeled the continuum using the total \textit{Model 1}, we obtained the opposite, that is, a higher cumulative probability for the \textsc{diskline} model. Therefore, we conclude that for the observations analyzed in this paper the broad emission Fe line profiles can be described as a signature of a wind or outflow and also by GR effects in terms of $\chi^2$-statistic and run-test.

The statistical run-test may allow a better assessment of the goodness of the nonrelativistic (\textsc{windline}) versus the relativistic (\textsc{diskline}) line profiles. To break the degeneracy between the relativistic and nonrelativistic broad iron lines, a study considering several other NS LMXB sources, containing strong and broad iron emission lines in their spectra could be performed. It could be performed in a perfect scenario, considering mainly sources with a more straightforward total spectrum, and using observations that are not affected by pile-up.

In the \textsc{windline} model, the large amount of mass outflowing with high velocity implies that the Earth observer should see broad and redshifted iron lines formed in the outflow. The velocity of the outflows in NS LMXBs given by the model could, for example, be compared with velocities found by P Cygni profiles in optical, UV, and X-ray because these lines indicate outflows or disk-wind outflows \citep[e.g.,][]{Brandt2000,Schulz2002}. If the \textsc{windline} model is found to be more appropriate to fit broad and skewed iron lines in NS LMXBs, the observation of such profiles will be of utmost importance for understanding outflowing in these systems; this is also an important tool for studying the inflowing-outflowing connection, as stated by \citet{Trigo2012}.

We emphasize that in addition to the moderate spectral resolution of the X-ray detectors that are available today, a careful analysis of the Fe K emission line can bring important information for understanding the physics behind the asymmetric Fe fluorescent emission lines and their implication on the neutron star physics. The next generation of X-ray observatories loaded with instruments with high spectral resolution will certainly improve the understanding of the asymmetric Fe line formation in accreting compact objects. 

We conclude this paper by also pointing out the importance of a timing variability study to constrain the physical process that leads to the broad and red-skewed iron line profiles. It can provide important and additional information for breaking the ambiguity between the relativistic and nonrelativistic line models.

\section*{Acknowledgments}
T. Maiolino acknowledges the financial support given by the Erasmus Mundus Joint Doctorate Program by Grants Number 2013-1471 from the agency EACEA of the European Commission, the CNES/INTEGRAL, and the CNR/INAF-IASF Bologna. T. Maiolino would also like to thank Cl\'{e}ment Stahl and Lorella Angelini for their valuable comments that contributed to this manuscript. We would like to thank Maria D\'{i}az Trigo for sharing details of previous XMM-Newton Epic pn data reprocessing. Finally, we thank the anonymous referee for their critical comments that considerably improved the content of the paper.

\bibliographystyle{aa}
\bibliography{bibliografiaarxiv}

\begin{thebibliography}{55}
\expandafter\ifx\csname natexlab\endcsname\relax\def\natexlab#1{#1}\fi

\bibitem[{{Allen} {et~al.}(2016){Allen}, {Schulz}, {Homan}, \&
  {Chakrabarty}}]{Allen2016}
{Allen}, J., {Schulz}, N.~S., {Homan}, J., \& {Chakrabarty}, D. 2016, in
  {AAS/High Energy Astrophysics Division}, Vol.~15, {AAS/High Energy
  Astrophysics Division}, 120.01

\bibitem[{{Arnaud}(1996)}]{Arnaud1996}
{Arnaud}, K.~A. 1996, in Astronomical Society of the Pacific Conference Series,
  Vol. 101, Astronomical Data Analysis Software and Systems V, ed. G.~H.
  {Jacoby} \& J.~{Barnes}, 17

\bibitem[{{Barlow}(1989)}]{Barlow1989}
{Barlow}, R. 1989, {Statistics. A guide to the use of statistical methods in
  the physical sciences} (John Wiley \& Sons Ltd)

\bibitem[{{Bhattacharyya} \& {Strohmayer}(2007)}]{Bhattacharyya2007}
{Bhattacharyya}, S. \& {Strohmayer}, T.~E. 2007, \apjl, 664, L103

\bibitem[{{Boirin} {et~al.}(2005){Boirin}, {M{\'e}ndez}, {D{\'{\i}}az Trigo},
  {Parmar}, \& {Kaastra}}]{Boirin2005}
{Boirin}, L., {M{\'e}ndez}, M., {D{\'{\i}}az Trigo}, M., {Parmar}, A., \&
  {Kaastra}, J. 2005, in SF2A-2005: Semaine de l'Astrophysique Francaise, ed.
  F.~{Casoli}, T.~{Contini}, J.~M. {Hameury}, \& L.~{Pagani}, 443

\bibitem[{{Brandt} \& {Schulz}(2000)}]{Brandt2000}
{Brandt}, W.~N. \& {Schulz}, N.~S. 2000, \apjl, 544, L123

\bibitem[{{Cackett} \& {Miller}(2013)}]{Cackett2013}
{Cackett}, E.~M. \& {Miller}, J.~M. 2013, \apj, 777, 47

\bibitem[{{Cackett} {et~al.}(2010){Cackett}, {Miller}, {Ballantyne}, {Barret},
  {Bhattacharyya}, {Boutelier}, {Miller}, {Strohmayer}, \&
  {Wijnands}}]{Cackett2010}
{Cackett}, E.~M., {Miller}, J.~M., {Ballantyne}, D.~R., {et~al.} 2010, \apj,
  720, 205

\bibitem[{{Cackett} {et~al.}(2008){Cackett}, {Miller}, {Bhattacharyya},
  {Grindlay}, {Homan}, {van der Klis}, {Miller}, {Strohmayer}, \&
  {Wijnands}}]{Cackett2008}
{Cackett}, E.~M., {Miller}, J.~M., {Bhattacharyya}, S., {et~al.} 2008, \apj,
  674, 415

\bibitem[{{Chiang} \& {Fabian}(2011)}]{Chiang2011}
{Chiang}, C.-Y. \& {Fabian}, A.~C. 2011, \mnras, 414, 2345

\bibitem[{{D'A{\`\i}} {et~al.}(2014){D'A{\`\i}}, {Iaria}, {Di Salvo}, {Riggio},
  {Burderi}, \& {Robba}}]{Dai2014}
{D'A{\`\i}}, A., {Iaria}, R., {Di Salvo}, T., {et~al.} 2014, \aap, 564, A62

\bibitem[{{Dauser} {et~al.}(2010){Dauser}, {Wilms}, {Reynolds}, \&
  {Brenneman}}]{Dauser2010}
{Dauser}, T., {Wilms}, J., {Reynolds}, C.~S., \& {Brenneman}, L.~W. 2010,
  \mnras, 409, 1534

\bibitem[{{D{\'{\i}}az Trigo} {et~al.}(2012){D{\'{\i}}az Trigo}, {Sidoli},
  {Boirin}, \& {Parmar}}]{Trigo2012}
{D{\'{\i}}az Trigo}, M., {Sidoli}, L., {Boirin}, L., \& {Parmar}, A.~N. 2012,
  \aap, 543, A50

\bibitem[{{Eadie} {et~al.}(1971){Eadie}, {Drijard}, \& {James}}]{Eadie1971}
{Eadie}, W.~T., {Drijard}, D., \& {James}, F.~E. 1971, {Statistical methods in
  experimental physics} (North-Holland Pub. Co.)

\bibitem[{{Fabian} {et~al.}(2000){Fabian}, {Iwasawa}, {Reynolds}, \&
  {Young}}]{Fabian2000}
{Fabian}, A.~C., {Iwasawa}, K., {Reynolds}, C.~S., \& {Young}, A.~J. 2000,
  \pasp, 112, 1145

\bibitem[{{Fabian} {et~al.}(1989){Fabian}, {Rees}, {Stella}, \&
  {White}}]{Fabian1989}
{Fabian}, A.~C., {Rees}, M.~J., {Stella}, L., \& {White}, N.~E. 1989, \mnras,
  238, 729

\bibitem[{{Fleischman}(1985)}]{Fleischman1985}
{Fleischman}, J.~R. 1985, \aap, 153, 106

\bibitem[{{Fridriksson} {et~al.}(2015){Fridriksson}, {Homan}, \&
  {Remillard}}]{Fridriksson2015}
{Fridriksson}, J.~K., {Homan}, J., \& {Remillard}, R.~A. 2015, \apj, 809, 52

\bibitem[{{Garc{\'{\i}}a} {et~al.}(2014){Garc{\'{\i}}a}, {Dauser}, {Lohfink},
  {Kallman}, {Steiner}, {McClintock}, {Brenneman}, {Wilms}, {Eikmann},
  {Reynolds}, \& {Tombesi}}]{Garzia2014}
{Garc{\'{\i}}a}, J., {Dauser}, T., {Lohfink}, A., {et~al.} 2014, \apj, 782, 76

\bibitem[{{Hagino} {et~al.}(2016){Hagino}, {Odaka}, {Done}, {Tomaru},
  {Watanabe}, \& {Takahashi}}]{Hagino2016}
{Hagino}, K., {Odaka}, H., {Done}, C., {et~al.} 2016, \mnras, 461, 3954

\bibitem[{{Hellier} \& {Mukai}(2004)}]{Hellier2004}
{Hellier}, C. \& {Mukai}, K. 2004, \mnras, 352, 1037

\bibitem[{{Laming} \& {Titarchuk}(2004)}]{Laming2004}
{Laming}, J.~M. \& {Titarchuk}, L. 2004, \apjl, 615, L121

\bibitem[{{Laor}(1991)}]{Laor1991}
{Laor}, A. 1991, \apj, 376, 90

\bibitem[{{Laurent} \& {Titarchuk}(2007)}]{Laurent2007}
{Laurent}, P. \& {Titarchuk}, L. 2007, \apj, 656, 1056

\bibitem[{{Lewin} \& {van der Klis}(2006)}]{Lewin2006}
{Lewin}, W.~H.~G. \& {van der Klis}, M. 2006, {Compact Stellar X-ray Sources},
  39 (Cambridge University Press)

\bibitem[{{Lewin} {et~al.}(1995){Lewin}, {van Paradijs}, \& {van den
  Heuvel}}]{Lewin1995}
{Lewin}, W.~H.~G., {van Paradijs}, J., \& {van den Heuvel}, E.~P.~J. 1995,
  {X-ray binaries} (Cambridge, Cambridge University Press), 252--307

\bibitem[{{Ludlam} {et~al.}(2017{\natexlab{a}}){Ludlam}, {Miller}, {Bachetti},
  {Barret}, {Bostrom}, {Cackett}, {Degenaar}, {Di Salvo}, {Natalucci},
  {Tomsick}, {Paerels}, \& {Parker}}]{Ludlam2017}
{Ludlam}, R.~M., {Miller}, J.~M., {Bachetti}, M., {et~al.} 2017{\natexlab{a}},
  \apj, 836, 140

\bibitem[{{Ludlam} {et~al.}(2017{\natexlab{b}}){Ludlam}, {Miller}, {Cackett},
  {Degenaar}, \& {Bostrom}}]{LudlamXTE2017}
{Ludlam}, R.~M., {Miller}, J.~M., {Cackett}, E.~M., {Degenaar}, N., \&
  {Bostrom}, A.~C. 2017{\natexlab{b}}, \apj, 838, 79

\bibitem[{{Lyu} {et~al.}(2014){Lyu}, {M{\'e}ndez}, {Sanna}, {Homan}, {Belloni},
  \& {Hiemstra}}]{Lyu2014}
{Lyu}, M., {M{\'e}ndez}, M., {Sanna}, A., {et~al.} 2014, \mnras, 440, 1165

\bibitem[{{Marinucci} {et~al.}(2014){Marinucci}, {Matt}, {Miniutti},
  {Guainazzi}, {Parker}, {Brenneman}, {Fabian}, {Kara}, {Arevalo},
  {Ballantyne}, {Boggs}, {Cappi}, {Christensen}, {Craig}, {Elvis}, {Hailey},
  {Harrison}, {Reynolds}, {Risaliti}, {Stern}, {Walton}, \&
  {Zhang}}]{Marinucci2014}
{Marinucci}, A., {Matt}, G., {Miniutti}, G., {et~al.} 2014, \apj, 787, 83

\bibitem[{{Matsuba} {et~al.}(1995){Matsuba}, {Dotani}, {Mitsuda}, {Asai},
  {Lewin}, {van Paradijs}, \& {van der Klis}}]{Matsuba1995}
{Matsuba}, E., {Dotani}, T., {Mitsuda}, K., {et~al.} 1995, \pasj, 47, 575

\bibitem[{{Miller}(2007)}]{Miller2007}
{Miller}, J.~M. 2007, \araa, 45, 441

\bibitem[{{Miller} {et~al.}(2002){Miller}, {Fabian}, {Wijnands}, {Remillard},
  {Wojdowski}, {Schulz}, {Di Matteo}, {Marshall}, {Canizares}, {Pooley}, \&
  {Lewin}}]{Miller2002b}
{Miller}, J.~M., {Fabian}, A.~C., {Wijnands}, R., {et~al.} 2002, \apj, 578, 348

\bibitem[{{Miller} {et~al.}(2011){Miller}, {Maitra}, {Cackett},
  {Bhattacharyya}, \& {Strohmayer}}]{Miller2011}
{Miller}, J.~M., {Maitra}, D., {Cackett}, E.~M., {Bhattacharyya}, S., \&
  {Strohmayer}, T.~E. 2011, \apjl, 731, L7

\bibitem[{{Miniutti} \& {Fabian}(2004)}]{Miniutti2004}
{Miniutti}, G. \& {Fabian}, A.~C. 2004, \mnras, 349, 1435

\bibitem[{{Mizumoto} {et~al.}(2018){Mizumoto}, {Done}, {Hagino}, {Ebisawa},
  {Tsujimoto}, \& {Odaka}}]{Mizumoto2018}
{Mizumoto}, M., {Done}, C., {Hagino}, K., {et~al.} 2018, \mnras

\bibitem[{{Nandra} {et~al.}(1997){Nandra}, {George}, {Mushotzky}, {Turner}, \&
  {Yaqoob}}]{Nandra1997}
{Nandra}, K., {George}, I.~M., {Mushotzky}, R.~F., {Turner}, T.~J., \&
  {Yaqoob}, T. 1997, \apjl, 488, L91

\bibitem[{{Orlandini} {et~al.}(2012){Orlandini}, {Frontera}, {Masetti},
  {Sguera}, \& {Sidoli}}]{Orlandini2012}
{Orlandini}, M., {Frontera}, F., {Masetti}, N., {Sguera}, V., \& {Sidoli}, L.
  2012, \apj, 748, 86

\bibitem[{{Pandel} {et~al.}(2008){Pandel}, {Kaaret}, \& {Corbel}}]{Pandel2008}
{Pandel}, D., {Kaaret}, P., \& {Corbel}, S. 2008, \apj, 688, 1288

\bibitem[{{Pintore} {et~al.}(2014){Pintore}, {Sanna}, {Di Salvo}, {Guainazzi},
  {D'A{\`\i}}, {Riggio}, {Burderi}, {Iaria}, \& {Robba}}]{pintore2014}
{Pintore}, F., {Sanna}, A., {Di Salvo}, T., {et~al.} 2014, \mnras, 445, 3745

\bibitem[{{Redman} \& {Rankin}(2009)}]{Redman2009}
{Redman}, S.~L. \& {Rankin}, J.~M. 2009, \mnras, 395, 1529

\bibitem[{{Reeves} {et~al.}(2006){Reeves}, {Fabian}, {Kataoka}, {Kunieda},
  {Markowitz}, {Miniutti}, {Okajima}, {Serlemitsos}, {Takahashi}, {Terashima},
  \& {Yaqoob}}]{Reeves2006}
{Reeves}, J.~N., {Fabian}, A.~C., {Kataoka}, J., {et~al.} 2006, Astronomische
  Nachrichten, 327, 1079

\bibitem[{{Reis} {et~al.}(2009){Reis}, {Fabian}, \& {Young}}]{Reis2009}
{Reis}, R.~C., {Fabian}, A.~C., \& {Young}, A.~J. 2009, \mnras, 399, L1

\bibitem[{{Reynolds} \& {Begelman}(1997)}]{Reynolds1997}
{Reynolds}, C.~S. \& {Begelman}, M.~C. 1997, \apj, 488, 109

\bibitem[{{Reynolds} \& {Nowak}(2003)}]{Reynolds2003}
{Reynolds}, C.~S. \& {Nowak}, M.~A. 2003, \physrep, 377, 389

\bibitem[{{Schulz} \& {Brandt}(2002)}]{Schulz2002}
{Schulz}, N.~S. \& {Brandt}, W.~N. 2002, \apj, 572, 971

\bibitem[{{Shaposhnikov} {et~al.}(2009){Shaposhnikov}, {Titarchuk}, \&
  {Laurent}}]{Shaposhnikov2009}
{Shaposhnikov}, N., {Titarchuk}, L., \& {Laurent}, P. 2009, \apj, 699, 1223

\bibitem[{{Sidoli} {et~al.}(2002){Sidoli}, {Parmar}, {Oosterbroek}, \&
  {Lumb}}]{Sidoli2002}
{Sidoli}, L., {Parmar}, A.~N., {Oosterbroek}, T., \& {Lumb}, D. 2002, \aap,
  385, 940

\bibitem[{{Soong} {et~al.}(1990){Soong}, {Gruber}, {Peterson}, \&
  {Rothschild}}]{Soong1990}
{Soong}, Y., {Gruber}, D.~E., {Peterson}, L.~E., \& {Rothschild}, R.~E. 1990,
  \apj, 348, 641

\bibitem[{{Tanaka} {et~al.}(1995){Tanaka}, {Nandra}, {Fabian}, {Inoue},
  {Otani}, {Dotani}, {Hayashida}, {Iwasawa}, {Kii}, {Kunieda}, {Makino}, \&
  {Matsuoka}}]{Tanaka1995b}
{Tanaka}, Y., {Nandra}, K., {Fabian}, A.~C., {et~al.} 1995, \nat, 375, 659

\bibitem[{{Titarchuk} {et~al.}(2009){Titarchuk}, {Laurent}, \&
  {Shaposhnikov}}]{Titarchuk2009}
{Titarchuk}, L., {Laurent}, P., \& {Shaposhnikov}, N. 2009, \apj, 700, 1831

\bibitem[{{Titarchuk} {et~al.}(2007){Titarchuk}, {Shaposhnikov}, \&
  {Arefiev}}]{TS2007}
{Titarchuk}, L., {Shaposhnikov}, N., \& {Arefiev}, V. 2007, \apj, 660, 556

\bibitem[{{Ueda} {et~al.}(2001){Ueda}, {Asai}, {Yamaoka}, {Dotani}, \&
  {Inoue}}]{Ueda2001}
{Ueda}, Y., {Asai}, K., {Yamaoka}, K., {Dotani}, T., \& {Inoue}, H. 2001,
  \apjl, 556, L87

\bibitem[{{Ueda} {et~al.}(2004){Ueda}, {Murakami}, {Yamaoka}, {Dotani}, \&
  {Ebisawa}}]{Ueda2004}
{Ueda}, Y., {Murakami}, H., {Yamaoka}, K., {Dotani}, T., \& {Ebisawa}, K. 2004,
  \apj, 609, 325

\bibitem[{{Vrielmann} {et~al.}(2005){Vrielmann}, {Ness}, \&
  {Schmitt}}]{Vrielmann2005}
{Vrielmann}, S., {Ness}, J.-U., \& {Schmitt}, J.~H.~M.~M. 2005, \aap, 439, 287

\end{thebibliography}

\end{document}